\documentclass{aa}

\usepackage{graphicx}
\usepackage{amsmath}
\usepackage{txfonts}
\DeclareUnicodeCharacter{2212}{-}

\begin{document}

   \title{Dust and gas content of high-redshift galaxies hosting obscured AGN in the CDF-S}

  % \subtitle{I. Overviewing the $\kappa$-mechanism}

   \author{Q. D'Amato \inst{1,2}\thanks{\email{quirino.damato2@unibo.it}}
          \and
          R. Gilli\inst{3}
          \and
          C. Vignali\inst{2,3}
          \and
          M. Massardi\inst{4}
          \and
          F. Pozzi\inst{2}
          \and
          G. Zamorani\inst{3}
          \and
          C. Circosta\inst{5}
          \and
          F. Vito\inst{6,7}
          \and
          J. Fritz\inst{8}
          \and
          G. Cresci\inst{9}
          \and
          V. Casasola\inst{1}
          \and
          F. Calura\inst{3}
          \and
          A. Feltre\inst{10}
          \and
          V. Manieri\inst{5}
          \and
          D. Rigopoulou\inst{11}
          \and
          P. Tozzi\inst{9}
          \and
          C. Norman\inst{12,13}
          }

   \institute{INAF/IRA, Istituto di Radioastronomia, Via Piero Gobetti 101, 40129, Bologna, Italy
         \and
             Dipartimento di Fisica e Astronomia dell’Università degli Studi di Bologna, via P. Gobetti 93/2, 40129 Bologna, Italy
         \and  
             INAF/OAS, Osservatorio di Astrofisica e Scienza dello Spazio di Bologna, via P. Gobetti 93/3, 40129 Bologna, Italy
         \and  
             INAF, Istituto di Radioastronomia - Italian Alma Regional Center (ARC), Via Piero Gobetti 101, 40129, Bologna, Italy
         \and     
             European Southern Observatory, Karl-Schwarzschild-Str 2, D-85748 Garching bei München, Germany
         \and     
             Instituto de Astrofísica and Centro de Astroingenieria, Pontificia Universidad Católica de Chile, Casilla 306, Santiago 22, Chile
         \and     
             Chinese Academy of Sciences South America Center for Astronomy, National Astronomical Observatories, Beijing 100012, China
         \and
             Instituto de Radioastronomía y Astrofisíca, UNAM, Campus Morelia, A.P. 3-72, C.P. 58089, Mexico
         \and
             INAF - Osservatorio Astrofisico di Arcetri, Largo E. Fermi 5, I-50125 Firenze, Italy
         \and
             SISSA, Via Bonomea 265, I-34136 Trieste, Italy
         \and
             Department of Physics, University of Oxford, Keble Road, Oxford OX1 3RH, UK
         \and
             Department of Physics and Astronomy, Johns Hopkins University, Baltimore, MD 21218, USA
         \and
             Space Telescope Science Institute, 3700 San Martin Drive, Baltimore, MD 21218, USA
         }

\date{Received XXX; accepted XXX}

% \abstract{}{}{}{}{} 
% 5 {} token are mandatory
 
  \abstract
  % context heading (optional)
  % {} leave it empty if necessary  
   {Obscured active galactic nuclei (AGN) represent a significant fraction of the entire AGN population, especially at high redshift ($\sim$70\% at z=3--5). They are often characterized by the presence of large gas and dust reservoirs that are thought to sustain and possibly obscure vigorous star formation processes that make these objects shine at far-IR and submillimeter wavelengths. Studying the physical properties of obscured AGN and their host galaxies is crucial to shedding light on the early stages of a massive system lifetime.
   }
  % aims heading (mandatory)
   {We aim to investigate the contribution of the interstellar medium (ISM) to the obscuration of quasars in a sample of distant highly star forming galaxies and to unveil their morphological and kinematics properties.
   }
  % methods heading (mandatory)
   {We exploit Atacama Large Millimeter/submillimeter Array (ALMA) Cycle 4 observations of the continuum ($\sim$2.1mm) and high-J CO emission of a sample of six X-ray selected, far-IR detected galaxies hosting an obscured AGN at $\mathrm{z_{spec}}>2.5$ in the 7 Ms \emph{Chandra} Deep Field-South (CDF-S). We measured the masses and sizes of the dust and molecular gas by fitting the images, visibilities, and spectra, and we derived the gas density and column density on the basis of a uniform sphere geometry. Finally, we compared the measured column densities with those derived from the Chandra X-ray spectra.
   }
  % results heading (mandatory)
   {We detected both the continuum and line emission for three sources for which we measured both the flux density and size. For the undetected sources, we derived an upper limit on the flux density from the root mean square (rms) of the images.
  %{We detected 3 sources out of six in both continuum and CO. For the undetected sources we obtained upper limits.
   We found that the detected galaxies are rich in gas and dust (molecular gas mass in the range <$0.5$ - $2.7 \times 10^{10}~M_\odot$ for $\alpha_{CO}=0.8$ and up to $\sim 2 \times 10^{11}~M_\odot$ for $\alpha_{CO}=6.5$, and dust mass <$0.9$ - $4.9 \times 10^{8}~M_\odot$) and generally compact (gas major axis 2.1-3.0 kpc, dust major axis 1.4-2.7 kpc). The column densities associated with the ISM are on the order of $10^{23-24}~\mathrm{cm^{-2}}$, which is comparable with those derived from the X-ray spectra. For the detected sources we also derived dynamical masses in the range $0.8$ -  $3.7 \times 10^{10}~M_\odot$.
}%
  % conclusions heading (optional), leave it empty if necessary 
   {We conclude that the ISM of high redshift galaxies can substantially contribute to nuclear obscuration up to the Compton-thick ($> 10^{24}~\mathrm{cm^{-2}}$) regime. In addition, we found that all the detected sources show a velocity gradient reminding one rotating system, even though two of them show peculiar features in their morphology that can be associated with a chaotic, possibly merging, structure.}
   

   \maketitle
%
%-------------------------------------------------------------------

\section{Introduction}
Active galactic nuclei (AGN) are characterized by accretion onto super massive black holes (SMBHs) which powers their emission over twenty decades in frequency. Almost all massive galaxies are thought to host a central SMBH which has passed (and possibly will pass) through active accretion phases during their life. Indeed, there is increasing evidence for a close connection between the star formation rate (SFR) and BH accretion history, suggesting a coevolution of the BH and its host galaxy; in particular, a crucial phase of such evolution occurs at $z$ = 2 -- 3, when both the SFR and BH accretion reach their maximum over cosmic time \cite[e.g.,][]{gruppioni_2013,madau_2014}.
The unified models postulate that nuclear absorption is due to a parsec-scale medium distributed in a toroidal geometry around the central engine \citep{urry_1995}. Nevertheless, the large amount of gas required to sustain the vigorous star formation (SF) of these objects during the epoch of their activity peak suggests that the UV and X-ray emission from the very central region of the galaxy can be obscured by the interstellar medium (ISM) located up to kiloparsec-scale in the host galaxy.
In addition, the high-z ($3<$ $z$ $\lesssim5$) heavily obscured (column density $\log N_H > 23$) AGN fraction is estimated to be $\sim$70\% \citep{vito_2018}, which is significantly higher than in the local Universe \cite[intrinsic distribution $\sim$43\% for the same column density,][]{burlon_2011}. This suggests an evolution of the luminous obscured AGN fraction, which may be driven by the larger gas content observed in high-z galaxies \cite[e.g.,][]{carilli_2013}.

Several concurring physical processes are responsible for the formation of galaxies and the build-up of their mass: for example, major mergers between gas--rich galaxies \cite[e.g.,][]{hopkins_2006}, mass inflows from the cosmic web \citep{dekel_2009,bournaud_2011}, or fast collapse in early halos during which most of the mass of the system is accreted \cite[e.g.,][]{lapi_2014,lapi_2018}; in all these scenarios the central regions are fueled with gas, triggering SF and BH accretion. Recent data from ALMA, Institut de radioastronomie millimétrique (IRAM) and Herschel indicate short gas depletion timescales \cite[some $10^8$ yr, see][]{genzel_2015,tacconi_2018}. The submillimeter galaxies (SMGs, defined as submilllimeter objects having flux densities $\gtrsim$ 1 mJy at 850 $\mu$m, \citeauthor{carilli_2013} \citeyear{carilli_2013}; see also \citeauthor{circosta_2019} \citeyear{circosta_2019} and references therein) are often indicated as those objects that trace the brief phase at the peak of SFR in the galaxy evolution.
They are usually detected at median redshift $z$ $\sim $2-3 \citep{simpson_2014}, and are characterized by extremely dense ISM, showing molecular masses up to $10^{10}~M_{\odot}$ and dust masses up to few $\times 10^{8}~M_{\odot}$ distributed on kiloparsec scales \citep{swinbank_2014,barro_2014,ikarashi_2015,massardi_2018,lapi_2018}.
The estimated SFR of these objects is $\sim 100 - 1000~M_{\odot}~\mathrm{yr^{-1}}$ \citep{blain_2002,swinbank_2014}. These properties make SMGs the best targets to investigate the contribution of the host galaxy in obscuring the AGN. Moreover, many observations revealed that these sources have morphologies showing complex gas motions and that the star--forming gas can lie in compact, rotating disk structures \citep{tacconi_2008,hodge_2012,ivison_2013,debreuck_2014}. Given their high gas and dust density and small size, SMGs are also considered among the best candidates as progenitors of compact quiescent galaxies \cite[cQGs][]{cimatti_2008}, a class of quiescent galaxies that are considerably smaller and denser than local galaxy counterparts of similar mass, showing stellar half--light radii of $\sim 0.5 - 2$ kpc and typical stellar masses of $\sim 10^{10} M_\odot$ \citep{cassata_2011}. %About 1/3 of cQGs contain an AGN, whereas the non--compact star forming galaxies have n AGN fraction < 1\%, thus suggesting that AGN feedback plays a relevant role in the evolution of these systems \citep{zubovas_2013}. 

Estimating the column density $N_H$ associated with the ISM of the host galaxy requires a multiwavelength approach exploiting data from the X-rays to the submillimeter: on the one hand, the X-rays directly probe the innermost region of the sources and can be used to evaluate the line of sight absorption to the central engine emission. On the other hand, the mm band is crucial to estimate the ISM properties such as the mass and geometrical distribution of the gas and dust components.
In particular, measurement of the gas mass, dominated by the molecular component at $z$ $\gtrsim$ 2 \citep{carilli_2013}, can be achieved with different methods: the most common exploits the low-J transition of CO as a tracer of the total molecular gas mass. An alternative method exploits the monochromatic continuum dust luminosity at 850 $\mu$m ($L_{850\mu m}$): \cite{scoville_2016} showed an empirical relation between $L_{850\mu m}$ and the CO(1--0) emission line luminosity -- valid for local ultraluminous infrared galaxies (ULIRGs, $L_{8-1000 \mu m}$ $\gtrsim 10^{12}~L_{\odot}$), z$\sim$2 SMGs and local star-forming galaxies -- from which they derived a direct relation between $L_{850\mu m}$ and the molecular gas mass, assuming a Galactic CO-to-$H_2$ conversion factor. Once the gas mass and size are measured through high resolution observation, the gas column density can be derived assuming a geometrical configuration.

\cite{gilli_2014} found that the column densities derived from the X-ray spectral analysis and the ISM mass are comparable for a SMG at $z$ = 4.755 hosting a heavily obscured AGN (XID 539, included in our sample, Table \ref{tab:sample}). This source is also included in the sample studied by \cite{circosta_2019}, who performed an in-depth analysis on seven X-ray selected, far infrared (FIR) detected AGN with spectroscopic redshift $z$ > 2.5 in the 7 Ms \emph{Chandra} Deep Field-South (CDF-S), in order to assess whether and to what extent the host ISM contributes to the AGN obscuration. They performed X-ray spectral fitting to derive the column density responsible for the central obscuration, assuming different spectral models. Then they exploited the available UV-to-FIR photometric data to derive the 850 $\mu$m flux from spectral energy distribution (SED) fitting, and they converted it into gas mass assuming the $ L'_{1-0} \propto L_{850\mu m}$ \cite{scoville_2016} empirical relation and the CO-to-$H_2$ conversion factor $\alpha_{CO}$ = 0.8 $\mathrm{~M_{\odot}~(K~km~s^{-1}~pc^2)^{-1}}$. On the basis of a simple assumption of spherical gas distribution with uniform density, they derived column densities of the host ISM that are comparable with those extracted from the X-ray fitting, and concluded that the ISM likely plays a significant role in the obscuration.% However, these results lack measurements that directly trace both the mass and size of the gas component. 

In this work we analyze for the first time the millimetric emission of the high-J CO emission and the dust continuum of six heavily obscured AGN (X-ray column densities $N_{H,X}$ > $10^{23} \mathrm{cm^{-2}}$) in the CDF-S at spectroscopic $z$ > 2.5, included in the sample of \cite{circosta_2019} and observed with ALMA. In Sec. \ref{sec:samp_sel} we illustrate the criteria adopted to select the targets and summarize the sample properties. In Sec. \ref{sec:obs} we describe the ALMA observations and the data reduction process. In Sec. \ref{sec:data_an} we perform the imaging and obtain the flux density, the angular size and line shape of the sources from both the spectrum and visibility fitting. We exploit these quantities in Sec. \ref{sec:results} to derive the mass and physical size of both the dust and molecular gas content. As for the gas mass, we explore several cases with different assumptions on the involved conversion factors, to address the large uncertainties in deriving such mass from both high-J CO transitions and dust long-wavelength emission. Then we derive the column densities for a simple spherical geometry and compare them with those measured from the X-ray spectra by \cite{circosta_2019}. We also calculate the dust mass in the optically--thin regime and the dynamical mass of the two sources for which the rotating disk model has been tested. Finally, in Sec. \ref{sec:conclusions} we summarize and discuss the results and future perspectives.

%We also measure the gas and dust component sizes, and we exploit these measurements to evaluate,  (i.e. a sphere for all the sources and a rotating disk for two sources), the column densities relative to the ISM in our targets. We finally compare the obtained column densities with those derived by \cite{circosta_2019}.

%This paper is structured as follows:   We also discuss the morphology and peculiar properties of some of the objects. 

Throughout the paper we adopt a standard $\Lambda$CDM cosmology: $H_0 = 69.6$, $\Omega_M = 0.286$, $\Omega_\Lambda = 0.714$ \citep{bennett_2014}.
As for the reported velocities we adopt the radio definition v = $(\nu_0-\nu)/\nu_0$, where $\nu_0$ and $\nu$ are the rest and observed frequency, respectively, in a Barycentric (i.e., referred to the Sun and Earth center of mass) rest-frame.

\section{Sample selection}
\label{sec:samp_sel}

The Chandra Deep Field South \citep[CDF-S, $\sim 480$  $\mathrm{arcmin}^2$,][]{luo_2017} provides the best X-ray spectral information currently available for distant AGN, thanks to its 7-Ms exposure time \cite[the average flux limit within 1 arcmin from the center of the field is $F_{0.5-7 keV} \approx 2 \times 10^{-17}~ \mathrm{erg~s^{-1}~cm^{-2}} $,][]{luo_2017}. In addition, this field features an exceptional multiwavelength coverage that allows the characterization of the overall SED of most X-ray sources with extraordinary accuracy. In particular, deep IR observations have been carried out with the Hubble Space Telescope (HST), \textit{Herschel} and \textit{Spitzer} observatories. Moreover, many observations in the submillimeter band were taken with instruments such as the Submillimeter Common-User Bolometer Array (SCUBA) on the James Clerk Maxwell Telescope and the Large Apex BOlometer CAmera (LABOCA) on the Apex Telescope. The central area of the CDF--S (about 1/3 of the whole field) is covered by the Great Observatories Origins Deep Survey \cite[GOODS--S,][]{giavalisco_2004}, in which the Cosmic Near--IR Deep Extragalactic Legacy Survey \cite[CANDELS,][]{grogin_2011}, a powerful imaging survey of the distant universe, has been carried out with HST. CANDELS consists of two different areas and depth portions, the deeper and smaller CANDELS/Deep and the shallower and more extended CANDELS/Wide. The CDF-S was also observed by the Galaxy Evolution from Morphology and SEDs \cite[GEMS,][]{rix_2004} survey, a large-area (800 $\mathrm{arcmin^2}$) two-color (F606W and F850LP) imaging survey with the Advanced Camera for Surveys (ACS) on the HST. 

In order to select our sources, we combined the sample from \cite{vito_2013}, consisting in 34 X-ray selected AGN at $z>3$ in the 4-Ms CDF-S with eight heavily obscured quasars (QSOs) in the redshift range 1.1--3.7, X-ray selected in the 1--Ms CDF-S exposure and observed at 850 $\mu$m with SCUBA \citep{rigopoulou_2009}. To select our targets from this combined sample we applied the following criteria. Firstly we required a secure (errors < 300 km/s) spectroscopic redshift > 2.5 as given by \cite{luo_2017}, in order to avoid the large uncertainties related to the photometric redshift that would propagated in the derived fundamental quantities for our study (such as the intrinsic luminosity) and properly select the CO emission lines to target in the ALMA observations. The redshift threshold at $z$ = 2.5 was chosen as a compromise between a sufficiently large gas content in the galaxies (expected to be larger at high-z, such that the host galaxy obscuration might be important) and the size of the sample. Secondly we required at least one > 3$\sigma$ detection at $\lambda_{obs}$ > 100 $\mu$m, in order to characterize the FIR-SED and maximize the detection probability with ALMA. Finally we included among the sample only the sources with gas column density $\log N_H > 23$ (derived from the X-ray spectral fitting), in order to consider only heavily obscured objects and trace the evolutionary stage in which the AGN is still deeply buried and the SF is high.

This selection returns six targets spanning the redshift range $\sim$ 2.5 -- 4.8. This sample includes six of the seven sources studied by \cite{circosta_2019}, who adopted our same selection criteria except for the requirement of heavy obscuration. As pointed out by those authors, the fact that the FIR detection criterion is sufficient to select only obscured AGN indicates a connection between the SF-heated dust in the ISM that is responsible for the FIR to (sub)millimeter emission and the nuclear absorption. Despite the small number of sources (owing to the stringent selection criteria), this sample represents a complete and unique collection of objects: they are characterized by a relatively high number of photons (from $\sim$260 to $\sim$2000 in the 7--Ms exposure) in the observed 0.5-7 keV band which allows an accurate estimate of the X-ray absorption and by good IR coverage.
The IDs, coordinates and redshift are reported in Table \ref{tab:sample}.
%We also report the SFR \textbf{(forse questo non serve)}, the stellar mass \textbf{(questo lo riporto per confrontarlo poi con le masse dinamiche)} and the ISM column density derived by \cite{circosta_2019} from the SED fitting, along with the column density extracted from the X-ray spectral fitting. \textbf{(forse non ha senso qui dilungarsi sui modelli usati e sul fatto che alla fine si riporta il trasmission model....)}. In Sec. \ref{sec:results} we compare these results with those derived from our data.
Throughout the paper we adopt for our targets the \cite{luo_2017} X-ray Identification numbers (XIDs).

\begin{table}
\caption{\label{tab:sample} Summary of the sample.
}
\centering
\resizebox{\hsize}{!}{
\begin{tabular}{cccccc}
\hline\hline

XID & XID & CID & RA & DEC  & $z_{\mathrm{spec}}$ \\ %& SFR & $M_{\ast}$ & $N_{H,~ISM_{SED}}$ & $N_{H,~X}$ \\
(1) & (2) & (3) & (4) & (5) & (6) \\ % & (7) & (8) & (9) & (10) \\
\hline
34   & 42 & 326\tablefootmark{a} & $03^h 31^m 51^s.95$ & $-27^{\circ} 53'27''.2$ & 2.940\tablefootmark{b} \\
262 & 337 & 5479 & $03^h 32^m 18^s.85$ & $-27^{\circ} 51' 35''.7$ & 3.660 \\
403 & 539 & 273  & $03^h 32^m 29^s.27$ & $-27^{\circ} 56' 19''.8$ & 4.755 \\
412 & 551 & 6294 & $03^h 32^m 29^s.86$ & $-27^{\circ} 51' 06''.1$ & 3.700 \\
490 & 666 & 9834 & $03^h 32^m 35^s.72$ & $-27^{\circ} 49' 16''.4$ & 2.578 \\
546
& 746 & 10578 & $03^h 32^m 39^s.68$ & $-27^{\circ} 48' 51''.1$ & 3.064\\

\hline
\end{tabular}
}
\tablefoot{
(1) X-ray ID in \cite{xue_2011} catalog (4--Ms).
(2) X-ray ID in \cite{luo_2017} catalog (7--Ms).
(3) CANDELS ID in \cite{grogin_2011} catalog.
(4) J2000 right ascension and (5) declination.
(6) Spectroscopic redshift as given by \cite{luo_2017}.\\
\tablefoottext{a}{This source is not included in CANDELS, thus the ID from GEMS \citep{haussler_2007} is reported.}\\
\tablefoottext{b}{Using the CIV, HeII and NV lines in the UV--spectrum we derive a better estimate of $z_{spec}=2.937$ (See Sec. \ref{sec:data_an}).}
}
\end{table}

The selected objects are all included in CANDELS/Deep area, except for XID 539 that lies in CANDELS/Wide and XID 42 that is outside CANDELS, but it is included in the GEMS catalog, in the multiwavelength survey by Yale--Chile (MUSYC) catalog \cite[consisting of deep optical medium-band and broadband optical and near--IR imaging of the CFDF--S,][]{gawiser_2003}, and it has been also observed in the $H_{160}$ band by HST wide field camera 3 (WFC3) \citep{chen_2015}.

\citeauthor{circosta_2019} (\citeyear{circosta_2019}, see Table 4) derived their basic quantities (such as SFR, IR luminosity and stellar mass) by the SED decomposition technique, finding that the targets are massive (stellar mass $\sim 10^{11}~M_{\odot}$) and very luminous (infrared luminosity in the 8-1000 $\mu$m band $\gtrsim 10^{12}~ \mathrm{L_\odot/yr}$) galaxies. They derived SFRs of $\sim 200 - 900 ~\mathrm{M_\odot/yr}$ (except for XID 42 which has $\sim 1680~ \mathrm{M_\odot/yr}$), which places all sources within 1$\sigma$ from the best fit main sequence relation found by \cite{schreiber_2015} at the corresponding redshift (2$\sigma$ for XID 337 which is below the main sequence). Three of them are well known heavily obscured AGN: XID 337 was firstly detected in the early 1-Ms CDF-S exposure and classified as a type II QSO object in the spectroscopic follow--up \citep{szokoly_2004}. This result was later confirmed by \cite{mainieri_2005} who also reported a submillimeter detection at 850 $\mu$m using SCUBA. Exploiting high resolution ($\sim$0.2") ALMA observations, \cite{circosta_2019} found that the single object associated with XID 337 in the Spitzer/MIPS, Herschel and SCUBA maps is actually a blend of two sources, whose emission is dominated by a bright QSO at 3.5" separation; thus, the Mid-IR/Far-IR photometric datapoints of XID 337 were converted into upper limits. XID 551 was the first type II AGN ever detected in the CDF-S 1-Ms exposure at z>3 \citep{norman_2002}, for which \cite{comastri_2011} found a large column density ($\log N_H > 23$) performing an X-ray spectrum analysis using 3.3--Ms XMM-Newton survey data. XID 539 \cite[ALESS73.1,][]{hodge_2013} is the most distant Compton-thick AGN known \citep{gilli_2011}, and one of the widely studied sources in the CDF-S \citep{coppin_2010,nagao_2012,debreuck_2014}. In particular, \cite{gilli_2014} have performed an in--depth analysis through an observation of the continuum at $\sim$1.3 mm, combining it with continuum observations from ALESS and from \cite{nagao_2012}. They built a full SED including Herschel data and obtained dust temperature and size and the SFR. XID 666 (BzK 4892) is a Compton-thick QSO studied by \cite{feruglio_2011,delmoro_2016,perna_2018}. XID 42 (ALESS57.1) has been recently studied by \cite{calistro_2018} an by \cite{rybak_2019}, who resolved the CO(3-2) and  [\ion{C}{ii}] emission, respectively. XID 746 is the less studied objects, despite being known as Type II QSO since the first 1--Ms exposure in the CDF-S \citep{szokoly_2004}.

\section{Observations}
\label{sec:obs}

ALMA dramatically improved our knowledge of SMGs, providing us with extremely deep and high-resolution images and spectra, which can be used to probe the dust mass and temperature of these objects, their star formation activity, and their gas mass.

We performed ALMA observations (Cycle 4, Band 4, ID: 2015.1.01205.S, PI: R. Gilli) of the continuum at $\sim$ 2.1 mm (142.78 GHz) and of one high-J CO transition for each of the targets, reported in Table \ref{tab:transitions} along with their rest- and observed-frame frequencies. The requested sensitivity was $\sim 180 \mu$Jy/beam for the line emission (considering a line width of 200 km/s) that corresponds to 20 $\mu$Jy/beam for the continuum.

\begin{table*}
\caption{\label{tab:transitions} Summary of the CO and continuum observations.
}
\centering
%\resizebox{\hsize}{!}{
\begin{tabular}{cccccccccc}
\hline\hline

XID & CO line & $\nu_{rest}$ & $\nu_{obs}$ & $\mathrm{beam_{CO}}$ & $\mathrm{beam_{cont}}$& $\mathrm{rms_{CO}}$& $\mathrm{rms_{cont}}$ \\ 	(1) & (2) & (3) & (4) & (5) & (6) & (7) & (8) \\
\hline
42	& (5-4) & 576.27 & 146.13 & 0.280 $\times$ 0.246 & 0.335 $\times$ 0.310 & 347   & 19\\
337	& (6-5) & 691.47 & 148.38 & 0.573 $\times$ 0.532  & 0.592 $\times$ 0.542 & 274 & 26 \\
539	& (7-6) & 806.65 & 140.16 & 0.302 $\times$ 0.283  & 0.317 $\times$ 0.301 & 195     & 21  \\
551 & (6-5) & 691.47 & 147.12 & 0.572 $\times$ 0.531  & 0.591 $\times$ 0.541 & 261 & 26 \\
666	& (4-3)	& 461.04 & 128.85 & 0.326 $\times$ 0.306 & 0.302 $\times$ 0.298 & 312/273\tablefootmark{a} & 20 \\
746	& (5-4)	& 576.27 & 141.80 & 0.571 $\times$ 0.524 & 0.565 $\times$ 0.540 & 280   &  23\\
\hline
\end{tabular}
%}
\tablefoot{(1) ID of the sources. (2) Observed CO transition. (3) Rest and (4) observed frequency of the line (in units of GHz). (5) beam size of the CO images and (6) beam size of the continuum images (in units of arcsec). (7) Root mean square (rms) of the CO images and (8) rms of the continuum images (in units of $\mu$Jy/beam).\\
The imaging parameters and procedure used to achieve the reported resolution and sensitivity are described in details in Sec. \ref{subsec:CO_im} for the CO observation and in Sec. \ref{subsec:dust_im} for the continuum.\\
\tablefoottext{a}{The two values correspond to the blue- and red-shifted peaks, respectively (see Fig. \ref{fig:spectra}).}\\
The rms is the defined as the mean rms measured on the images in four regions located at $\sim10$" from the sources.
}
\end{table*}

The observations were split in two blocks, each containing three targets: the $2 \times 4$ GHz spectral windows (\textit{spws}) cover the range 126.56-130.33 GHz and 138.56-142.33 GHz to observe XID 666, XID 746 and XID 539, and the range 133.07-136.93 GHz and 145.17-148.93 GHz for XID 42, XID 337 and XID 551. Each \textit{spw} is sampled in 240 channels of 7.812 MHz width corresponding, at the median frequency of the \textit{spws}, to 17.5 km/s in velocity for the first block and to 16.6 km/s for the second one.
The project was accepted in C priority and was only partially observed at the end of the Cycle 4. In particular, the first block was observed on the $23^{\mathrm{th}}$ September 2016 for about 88 min, and the second block for about 46 min on the $29^{\mathrm{th}}$ September 2016. QSO J0334-4008 was used to calibrate the flux and bandpass, while the QSO J0348-2749 served as phase calibrator\footnote{https://almascience.eso.org/alma-data/calibrator-catalogue}.

Three out of six sources (namely XID 42, XID 539 and XID 666) were detected in both the line and continuum emission. Concerning the undetected sources, two of them (XID 337 and XID 551) were observed for half of the time needed to achieve the requested sensitivity, while XID 746 is included in the execution block that was observed twice and reached the requested sensitivity.
%The original proposal of the observations of all the sample was made predicting the flux density in the ALMA band 4 ($\sim$2.1 mm) assuming a modified black--body model; this model was constrained on the basis of the available photometric data. Since the slope of this model in the Rayleigh--Jeans tail is a function of the  $\beta$ index for the optical depth ($S_\nu \propto \nu^{2+\beta}$), a slightly steeper index than assumed may explain why the source is not detected.

We performed data-reduction using the ALMA calibration pipeline in CASA version 4.7.0-1. The pipeline procedure automatically flags bad data and performs the calibration; it also produces several diagnostic plots that are useful to identify possible calibration failures and discern additional needed flags. After the inspection of both the diagnostic plots and calibrated data, the calibration resulted to be satisfactory.

\section{Data analysis}
\label{sec:data_an}

We performed a standard imaging of the sources by using the {\sc{clean}} task of CASA in ``velocity'' mode for the line emission and ``mfs'' mode for the continuum. The imaging parameters were chosen as best trade--off between sensitivity and angular resolution (see Sec. \ref{subsec:CO_im} and Sec. \ref{subsec:dust_im} for details, Table \ref{tab:transitions} for the resolution and sensitivity of each observation).

\begin{figure*}
\centering
		
\resizebox{\hsize}{!}{\includegraphics{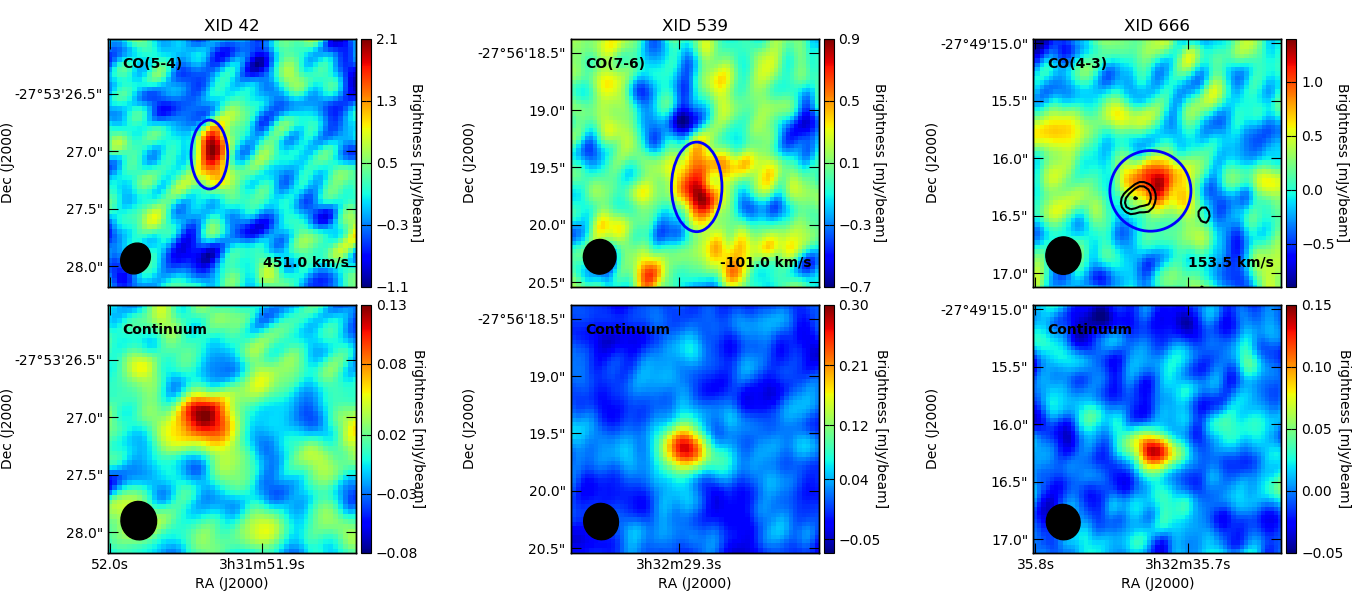}}

\caption{\emph{Top row}: CO-line peak channel of the three detected sources. The observed CO transition and the velocity corresponding to the center of the channel (defined as the velocity shift with respect to the systemic redshift) are reported in the up-left and bottom-right corner of the images, respectively. The blue ellipses are the regions from which the spectra in Fig. \ref{fig:spectra} were extracted. In the XID 666 CO-line peak image (right panel) we show the contours of the blue-shifted peak channel centered at -206.5 km/s ($2,3,4\sigma$, black solid lines) over the red-shifted peak channel (see Sec. \ref{sec:data_an} for details). \emph{Bottom row}: dust-continuum image of the three detected sources. In all the images the black ellipse in the left-bottom corner represents the clean beam.
}

\label{fig:imaging}

\end{figure*}

\subsection{CO--line imaging}
\label{subsec:CO_im}

Since the line is not clearly visible from the visibility amplitude--frequency plane, we cannot unambiguously assess the channels containing it, thus we cannot a priori exclude them when we image the continuum emission. Thus we decided to image separately the \textit{spw} containing the line and the other \textit{spws} for the continuum, in order to avoid any possible contamination.
To image the continuum-subtracted line we preliminarly ran the {\sc{uvcontsub}} task, that performs a polynomial fitting of the continuum level and then subtracts it from the data. We fit the continuum using only the \textit{spws} that do not contain the line, and subtract the resulting fit from the \textit{spw} in which the line is expected to be. 
For XID 337, XID 551, and XID 746 it was not possible to determine unambiguously a channel in which the brightness at the source position was significantly above the noise level. For these sources no detection was obtained even applying natural weighting, tapering (0.45") and a large channel width (120 km/s) to improve the chance of achieving a significant detection.
For XID 42, XID 539 and XID 666 imaging we used a robust weighting equal to 0.5, 1, 1, and a channel width of 40, 120 and 40 km/s, respectively. We clearly identified the channels corresponding to the peak of the emission, whose images are reported in the top row of Fig. \ref{fig:imaging}. We detect XID 42, XID 539 and XID 666 with a signal-to-noise ratio (S/N) equal to 6.2, 4.7 and 4.0/4.9, respectively. The S/N is defined as the ratio between the value of the brightest pixel and the root mean square (rms) derived averaging the rms measured in four regions placed at $\sim$10" from any source. Since for XID 666 we detected a double peaked line, the two values refer to the blue- and red-shifted peaks. As a confirmation of the detection, we note the central position of the peak in the images (offset with respect to expected source positions < 0.5", beam sizes in Table \ref{tab:transitions}) and the line feature in the spectra visible in Fig. \ref{fig:spectra}.

We note that only for XID 539 the velocity of the peak channel (reported on the image) is consistent with zero, being the channel central velocity $-$111.6 km/s and its width is 120 km/s.
XID 666 has two peaks shifted respectively by $\sim -$200 km/s and $\sim$ 150 km/s centered on the rest-frame velocity. This double-peaked line, also clearly visible from the spectrum (Fig. \ref{fig:spectra}), is interpreted as a signature of a rotating structure. This interpretation is also supported by the velocity map (Fig. \ref{fig:moments}) that shows a continuous gradient along one direction. However, in Sec. \ref{subsec:col_dens} we briefly explore the possibility that this source is actually a blend of two nonresolved, merging system.
The spectroscopic redshift of XID 42 (z = 2.940) was derived by \cite{szokoly_2004} from the FOcal Reducer and low dispersion Spectrograph (FORS) of the Very Large Telescope (VLT).
Re-analyzing the same spectrum using the high excitation lines CIV and NV, and the recombination line HeII (that is a robust redshift estimator) we found a better estimate $z$ = 2.9369 $\pm$ 0.0003, assumed as the rest--frame hereafter (against the value z=2.939 obtained using the Ly-$\alpha$ line, which is known to be challenging as exact redshift estimator due to the resonant scattering). From the CO(5--4) line we derived a redshift $z$ = 2.94355 $\pm$ 0.00001, which is shifted by $\sim$ 500 km/s with respect to the rest--frame velocity. We note that  \cite{calistro_2018} derived a redshift $z$ = 2.943 $\pm$ 0.002 from the CO(3--2) line, which is compatible within 3$\sigma$ with that derived both from the optical spectrum and from the CO(5--4) transition. This redshift is also consistent with the  [\ion{C}{ii}] line detected by \cite{rybak_2019}. The rest frequency displacement between the CO and [\ion{C}{ii}] lines and the UV emission lines could be explained as a difference in radial velocity. Such as spectral displacements have been often reported in the literature \citep{pentericci_2016,matthee_2017,carniani_2018}, and interpreted as a consequence of the clumpy structure of high-z galaxies, in which the different components move with respect to each other. High excitation UV-rest-frame lines and the CO (and  [\ion{C}{ii}]) lines indeed trace different gas phases and may arise from different regions.

\subsection{Moments}
\label{subsec:moments}
\begin{figure}
\resizebox{\hsize}{!}{\includegraphics{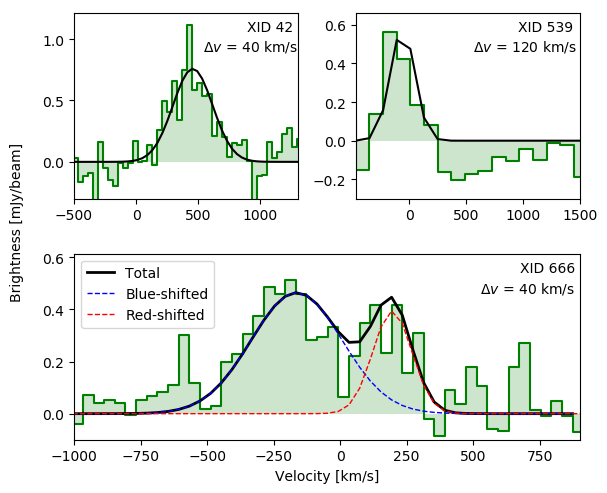}}
\caption{Spectrum of XID 42, XID 539 and XID 666 (green shaded histogram), expressed as the mean brightness within the regions contained in the blue ellipses of Fig. \ref{fig:imaging}. The black solid line represents the Gaussian line fitting. Due to its double-peaked feature, we fit the line of XID 666 with two Gaussian components, indicated in the plot with the blue and red dashed lines. The channel width $\Delta v$ is reported for each spectrum.}
\label{fig:spectra}
\end{figure}

By using the {\sc{immoments}} task we computed the moments 0, 1, 2 for the detected sources, defined as the frequency-integrated flux, the velocity field and the velocity dispersion, respectively. We performed a Gaussian fitting of the line spectrum and derived their central velocity $\mathrm{v_0}$ (with respect to the rest-frame velocity at the given redshift) and Full Width at Half Maximum (FWHM); to calculate the moments we considered a velocity range given by $\mathrm{v_0} \pm \mathrm{FWHM}$, in order to include few channels aside the line and avoid the loss of signal from its wings. For XID 666, to calculate the moments of the entire source, we consider the channels spanning the velocity range from $\mathrm{v_0} - \mathrm{FWHM}$ of the blue component to $\mathrm{v_0} + \mathrm{FWHM}$ of the red one. In Fig. \ref{fig:spectra} we show the spectra of the sources and their fits.

We fit the lines of XID 42 and XID 539 with a single Gaussian component, since there are not evident double-peaked features in their spectra. As for XID 666, we fit two Gaussian components on the whole source region (blue ellipse in Fig. \ref{fig:imaging}). Since the two spectral peaks of XID 666 are quite close and likely contaminated by each other, we additionally performed fitting with a single Gaussian component on the spectra of the two spatially separated regions corresponding to the peaks, in order to estimate such contamination and derive the most accurate line width. Both the line width and central velocity of the single blue and red shifted components are fully consistent (within $1\sigma$) with those derived from the spectrum over the entire source region. In the following, we use the line width and central velocity derived from the two Gaussian components fit on the whole region. The fit results are reported in Table \ref{tab:spectra_fit}. We note that the brightness of the peak is not used in any of the following considerations and derived physical quantities.

\begin{table}
\caption{\label{tab:spectra_fit} Results of the Gaussian line fitting performed on the spectra. 
}
\centering
%\resizebox{\hsize}{!}{
\begin{tabular}{cccc}
\hline\hline

XID & $B_{\mathrm{v_0}}$ & $\mathrm{v_0}$ & FWHM  \\
    & [mJy/beam] & [km/s] &[km/s]\\
\hline 
42           & $0.76 \pm 0.05$  & $498 \pm 14$  & $368 \pm 32$        \\
539	     & $0.47 \pm 0.06$  & $-56 \pm 33$  & $308 \pm 77$        \\       
$666_{Blue}$ & $0.46 \pm 0.02$  & $-194 \pm 26$  & $474 \pm 67$  \\
$666_{Red}$  & $0.39 \pm 0.03$  & $187 \pm 12$   & $162 \pm 27$  \\
\hline
\end{tabular}
%}
\tablefoot{Columns from left to right: source ID, brightness of the Gaussian peak, central velocity (with respect to the rest-frame at the given redshift) and FWHM. For XID 666 we report the two Gaussian components, simultaneously fit on the whole source region. The blue-shifted and red-shifted components (with respect to the rest-frame velocity) are named $666_{Blue}$ and $666_{Red}$, respectively.
}
\end{table}

\begin{figure*}[h!]
	\centering
\resizebox{\hsize}{!}{\includegraphics{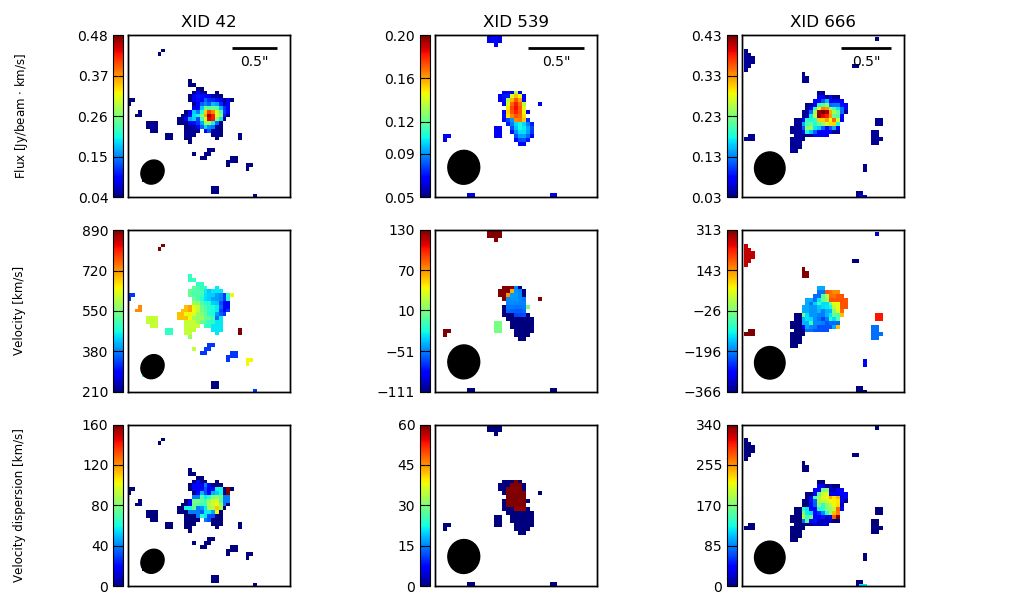}}

	\caption{Total flux (moment 0), velocity map (moment 1) and velocity dispersion map (moment 2) (from top to bottom) extracted for XID 42, XID 539 and XID 666 (from left to right). Only the pixels above the 3$\sigma$ level are shown. In the top right corner of the moment 0 images the angular scale is reported. In all the images the black ellipse in the left-bottom corner represents the clean beam.}
	
	\label{fig:moments}
\end{figure*}

We extracted the moments using as a threshold a value of $3\sigma$ calculated by the measured rms on the image. Moments are reported in Fig. \ref{fig:moments}.

\subsection{Dust continuum imaging}
\label{subsec:dust_im}

For each source we imaged all the \textit{spws} but those containing the CO line.
As for the line imaging, for XID 337, XID 551, and XID 746 we apply natural weighting and a 0.45" tapering, without achieving a significant detection. We detected the continuum emission for XID 42, XID 539 and XID 666 with a S/N equal to 6.8, 12.8 and 7.0, respectively, using a robust weighting equal to 1 for all of them. We show the continuum images of the detected sources in the bottom row of Fig. \ref{fig:imaging}.

\subsection{Visibility fitting}

Since the reconstructed images depend on the deconvolution algorithm and its input values, they do not provide a unique and direct representation of the data but rather a nonlinear mapping from the Fourier domain into the sky plane. Hence, we firstly perform an image fitting with the sole purpose of using the obtained values as starting input parameters to perform the fits directly on the visibilities, which are unaffected by the reconstruction algorithm and pixel correlation as images are. In addition, if the sources are not or barely spatially resolved in the images, the visibility fitting provides an estimate of their angular size even on scales smaller than the synthesized beam.

As for the image fitting, to derive the information about the dust component, we fit the continuum image, while for the gas we fit the moment 0 of the CO-line since it represents the total integrated emission. We performed the fitting using the task {\sc{imfit}}, which relies on the algorithm developed by \cite{condon_1997} using as a model a spatial elliptical Gaussian 2--D function. To select the region in which the fit is performed, for the continuum we chose a region relatively large ($\sim$1.5" diameter) around the sources to include all the signal, setting a  $3\sigma$ threshold on the pixels to include. Given the high S/N this selection guarantees that only the source will be fit. For moment 0 this is not an issue because all the included pixels are above the $3\sigma$ significance level and are attributed to the source. For the line fitting we included the same channels ($\mathrm{v_0} \pm \mathrm{FWHM}$) used to calculate the moments. %The fits returns the celestial and pixel coordinates of the Gaussian peak and its value, the major and minor axis FWHM of the clean beam, the major and minor axis FWHM of both the convoluted and deconvoluted Gaussian fit, the position angle, the spatially integrated flux density expressed in Jy for the continuum and the spatially integrated flux in $\mathrm{Jy\cdot km/s}$ for the moment 0.

Since the task {\sc{uvmodelfit}}, which performs the visibility fitting, takes the flux density as input, while {\sc{imfit}} returns the velocity-integrated flux, in order to obtain the starting value for the visibility fitting of the CO line emission we divided the integrated flux obtained from the image fitting by the FWHM of the line reported in Table \ref{tab:spectra_fit}. We included the same native channel range corresponding to $\mathrm{v_0} \pm \mathrm{FWHM}$ (on the continuum subtracted dataset) for the line fitting, and the rest of the band for the continuum. We initially left all the parameters free in order to check whether the central position of the Gaussian matches that derived from the image fitting. Since the offset was within $1\sigma$, we rerun a second fit keeping the position fixed in order to reduce the uncertainties related to the flux densities and the sizes. Final fitting results are reported in Table \ref{tab:results}.

\begin{table*}
\caption{\label{tab:results} Visibility fitting results CO line (top) and dust continuum (bottom) detections.
}
\centering
%\resizebox{\hsize}{!}{
\begin{tabular}{c  c  c  c  c  c  c}
%\rowcolors{1}{light-gray}{white}
\multicolumn{7}{c} {CO line} \\
\hline \hline
XID & Flux density & RA & DEC & Major axis & Axial ratio & Position angle  \\
 (1)   & (2)    &  (3)  &  (4)   &  (5) & (6)  &  (7) \\
\hline
%337  & -- & -- & -- & -- \\

%551  & -- & -- & -- & -- \\

42   & $1.5 \pm 0.1$ & $03^h 31^m 52^s.16$ $\pm 0^s.02$ & $-27^{\circ} 53'27''.32$ $\pm 0''.02$ & $0.38 \pm 0.04$ & $0.6 \pm 0.2$ &  $-35 \pm 13$ \\

539  & $0.7 \pm 0.1$ & $03^h 32^m 29^s.27$ $\pm 0^s.02$ & $-27^{\circ} 56' 19''.60$ $\pm 0''.05$ & $0.46 \pm 0.13$ & $0.6 \pm 0.3$ & $-46 \pm 30$ \\

%746  & -- & -- & -- & --  \\

666  & $1.01 \pm 0.07$ &  $03^h 32^m 35^s.60$ $\pm 0^s.01$ & $-27^{\circ} 49' 16''.16$ $\pm 0''.01$ & $0.26 \pm 0.04$ &$ 0.5 \pm 0.2$ & $-42 \pm 16$ \\
\hline
\end{tabular}

\begin{tabular}{c  c  c  c  c  c  c}
\multicolumn{7}{c} {} \\

\multicolumn{7}{c} {Dust Continuum} \\
\hline
\hline
XID & Flux density& RA & DEC & Major axis & axial ratio & Position angle  \\
 (1)   & (2)    &  (3)  &  (4)   &  (5) & (6)  &  (7) \\
\hline

%337 & -- & -- & -- & -- \\

%551 & -- & -- & -- & -- \\

42  & $0.23 \pm 0.02$ & $03^h 31^m 52^s.20$ $\pm 0^s.03$ & $-27^{\circ} 53'27''.28$ $\pm 0''.02$ & $0.34 \pm 0.07$ & $0.6 \pm 0.3$ & $ -75 \pm 23$\\

539 & $0.41 \pm 0.02$ & $03^h 32^m 29^s.36$ $\pm 0^s.01$ & $-27^{\circ} 56' 19''.63$ $\pm 0''.01$ & $0.27 \pm 0.03$ & $0.6 \pm 0.2$ & $70 \pm 13$\\

%746 & -- & -- & -- & -- \\

666 &  $0.19 \pm 0.02$ & $03^h 32^m 35^s.77$ $\pm 0^s.01$ & $-27^{\circ} 49' 16''.20$ $\pm 0''.01$  & $ 0.17 \pm 0.05$ & ... & $78 \pm 21$\\
\hline

\end{tabular}
%}
\tablefoot{
(1) ID of the sources. (2) Flux density at the peak of the 2-D Gaussian (in units of mJy). (3) RA and (4) DEC of the peak of the 2-D Gaussian function. (5) FWHM of the Gaussian major axis (in units of arcsec). (6) Ratio between the minor and the major axes. (7) Position angle (in units of degree), defined as the angle between the major axis of the fit and the northsouth axis of the field, ranging from -90$^\circ$ to 90$^\circ$, positive if clockwise.
}
\end{table*}

The large error associated with the position angle is due to the fact that the sources are barely resolved and it is difficult to unambiguously derive an orientation. However, this does not affect our analysis since we are interested in the size of the sources and not in their orientation with respect to the celestial axes.
For what concerns the dust, the continuum axial ratio ($R=b/a$, where $b$ is the minor axis and $a$ is the major axis) of XID 666 is consistent with 0. This is attributed to the very small size of the component that cannot be resolved in the minor axis direction.

\section{Results and discussion}
\label{sec:results}

\subsection{Source size}

Given the source redshifts, we derived the physical source size from the resulting fit. For XID 666, the continuum axial ratio could not be estimated either from the visibilities (Table \ref{tab:results}) or from the deconvoluted image fitting. However, from the images we obtained the size of $a$ and $b$ convoluted with the clean beam. Hence, for XID 666, a fair assumption on the value of $R$ is that derived from the convoluted $a$ and $b$, that is $R=0.8$. This ratio can be affected by the convolution given that, for Gaussian beams and sources, sizes add in quadrature. However, the derived ratio is consistent within $1\sigma$ with those derived for the other sources, and given the large uncertainties this assumption does not significantly affect our results. We do not propagate the errors on the axes as obtained from the image fitting to estimate the uncertainty in the continuum axial ratio of XID 666, since the error on $R$ derived from the visibility fitting is generally larger, as visible from the fits of the other sources in Table \ref{tab:results}. These (relative) errors span the range 30\%--50\%. Thus, using a conservative approach, we assigned a relative error of 50\% to $R$ for XID 666, corresponding to the maximum error obtained from the other fits. In Table \ref{tab:phys_dim} the physical sizes of each source are reported for both the molecular gas and dust components. In order to derive an upper limit of the gas density, for the undetected sources (XID 337, XID 551 and XID 746) we assume as ``typical'' major and minor axes the mean of those of the detected ones. Several analyses have shown that this kind of objects have all comparable sizes \cite[e.g.,][]{swinbank_2010,tacconi_2008}. In addition, the mean redshift $z=3.42$ of the detected sources is similar to that of the undetected ones ($z=3.47$), thus we may expect that they have a similar size. We still use a conservative approach assigning a 30\% relative error to $a$ and 50\% to $b$, corresponding to the maximum uncertainties obtained from the visibility fits. The assumed $a$ and $b$ of XID 337, XID 551, XID 746 are $2.7 \pm 0.8$ kpc  and  $ 1.6 \pm 0.8 $ kpc for the CO component, and $2.0 \pm 0.6$ kpc  and  $ 1.2\pm 0.6 $ kpc for the dust component, respectively.

We found that the sources are extremely compact objects, all having a major axis of the Gaussian fit $\lesssim $ 3 kpc. These results are in agreement with the hypothesis that these sources may be identified as the likely progenitors of the compact Quiescent Galaxies (cQGs), a class of extremely compact objects \cite[$\lesssim$ 1 kpc,][]{cimatti_2008,talia_2018} that have experienced a brief phase of intense SF, eventually quenched by the SF and AGN feedback. We note that the molecular gas size obtained from the CO(5--4) line fitting for XID 42 is nearly half of that derived by \cite{calistro_2018} on the basis of the CO(3--2) observation, and more similar to both the optical HST $H_{160}$ and  [\ion{C}{ii}] size, whose emission is dominated by a compact (radius $<$1 kpc) component \citep{rybak_2019}. The difference between the CO(3--2) and CO(5--4) region sizes could be attributed to the fact that the CO(5--4) emission traces the innermost, denser and more excited phase of the gas. Indeed, \cite{engel_2010} showed how the extent of the CO emission may decrease for increasing J-transitions. However, we cannot exclude that the size of the CO(5--4) component is biased by the sensitivity limit of the observation. As for XID 539, the dust size is in agreement with \cite{debreuck_2014}, who constrained the diameter of the 1 mm continuum emission to a region < 2 kpc. They also resolved the  [\ion{C}{ii}] line emission, which has been spatially fit with a Gaussian, obtaining a size of 2.4 $\pm$ 0.2 kpc half-light radius, corresponding to about twice the CO(7--6) radius detected in this work (defined as the half of the mean between the major and minor axes). However, we note that -- due to the large error on the CO(7--6) size -- the two measurements are still consistent within $\sim~2\sigma$.

We also found that the dust component is confined in a smaller region, ($\sim$ 70\% radius for XID 539 and XID 666, $\sim$ 90\% for XID 42) than the molecular gas component. Our results are in agreement with the typical size of the dust in SMGs that has been shown to be $\lesssim$ 2 kpc \cite[e.g.,][]{ikarashi_2015,hodge_2016}. \cite{calistro_2018} showed how the (apparent) compactness of the dust can be explained by a radial decrease of the gas column density and dust temperature toward the outskirts of the sources, on the basis of a radiative transfer model presented by \cite{weiss_2007}. It is interesting to note that \cite{casasola_2017}, analysing the gas and dust distribution of a sample of 18 nearby face-on galaxies, found that --on average-- the exponential scale-length of the dust-mass surface-density is $\sim$2.3 times higher than that of the molecular gas, that is the dust is distributed at larger radii than the molecular gas, contrary to what we found at high-z. However, we point out that given the sensitivity limit of our observation it is possible that we are tracing only the innermost, warmer phase of dust, whose emission flux increases linearly with the density and temperature (in the Rayleigh-Jeans regime, assuming a constant dust mass).

\begin{table}
\caption{\label{tab:phys_dim} Physical sizes of the detected XID sources.
}
\centering
%\resizebox{\hsize}{!}{
\begin{tabular}{ccc}

\multicolumn{3}{c}{Molecular gas}  \\
\hline 
\hline
XID & Major axis & Minor axis      \\
    &   [kpc]          &  [kpc]    \\
\hline                           
%337 & --               & --        \\
%551 & --               & --         \\
42  & $3.0 \pm 0.3$    & $1.8 \pm 0.6$  \\
539 & $3.0 \pm 0.9$    & $2.0 \pm 1.0$       \\
%746 & --               & --           \\
666 & $2.1 \pm 0.3$    & $1.1 \pm 0.5$  \\
%337, 551, 746 & $2.7 \pm 0.8$  &  $ 1.6 \pm 0.8 $ \\ 
\hline
\multicolumn{3}{c}{}\\   
\multicolumn{3}{c}{Dust}\\
\hline
\hline
XID & Major axis & Minor axis          \\
    &   [kpc]          &  [kpc]      \\
\hline                                   
%337 & --               & --           \\ 
%551 & --               & --           \\ 
42  & $2.7 \pm 0.5$    & $1.6 \pm 0.9$  \\   
539 & $1.8 \pm 0.2$    & $1.1 \pm 0.4$  \\ 
%746 & --               & --            \\
666 & $1.4 \pm 0.4$    & $1.1 \pm 0.7$  \\ 
%337, 551, 746 & $2.0 \pm 0.6$  &  $ 1.2\pm 0.6 $ \\  
\hline         
\end{tabular}
%}
\tablefoot{For XID 666 the minor axis of the dust component was derived from the convoluted major and minor axial ratio obtained from the image fit, as it was not possible to derive it from both the deconvoluted image fit and visibility fit.
}
\end{table}

\subsection{Dust mass}

\label{subsec:dust_mass}

We derive the mass of the dust $M_d$ from its emission in the optically thin regime:

$$M_d = \frac{D_L^2 S_{obs}}{k_\nu B_\nu(T_d) (1+z)}$$

where $z$ is the redshift, $D_L$ is the luminosity distance,  $S_{obs}$ is the flux density at the observed frequency $\nu_{obs}$, $B_\nu(T_d)$ is the Planck function of the dust emission at the temperature $T_d$ and $k_\nu$ is the dust opacity per mass unit ($\mathrm{g^{-1}~cm^{2}}$), assumed to scale with the frequency as $k_\nu =4[\nu/(1.2~\mathrm{THz })]^\beta$ \citep{draine_2007}. The index $\beta$ is set equal to 2.0 \cite[e.g.,][]{magnelli_2012,gilli_2014}. For the undetected sources, an upper limit on $S_{obs}$ is assumed equal to three times the rms measured on the continuum image. The rest-frame frequency $\nu=\nu_{obs}(1+z)$ is computed considering as $\nu_{obs}$ the median observed frequency of the spectral windows in which the imaging and fitting of the continuum have been performed. The temperatures $T_d$ of the dust have been derived from the SED fit with a single modified black body (MBB) component $S_\nu \propto B_\nu(T_{d}) (1-e^{-\tau_\nu})$ by means of the code developed by \cite{feltre_2012}, see also \cite{circosta_2019}. The fiducial error associated with such temperatures is on the order of $\approx \pm 5$ K. The photometric datapoints used in this work to perform the SED fitting and derive the dust temperatures are the same of the SEDs reported in \cite{circosta_2019}, taken from the UV-to mid-IR (MIR) provided by \cite{hsu_2014}, complemented with the FIR data from \emph{Herchel}/PACS and SPIRE and reported in the \cite{magnelli_2013} and \cite{oliver_2012} catalogs, respectively. Furthermore, we also added the measurements reported in this work at 2.1 mm. In addition to this, the other main difference with respect to \cite{circosta_2019} is that they fit the FIR SED using several templates accounting for the stellar, star formation and AGN contributions, rather than a single MBB component. We discuss the consequences of these differences in Sec. \ref{subsec:gas_mass}. The temperatures and dust masses are reported in Table \ref{tab:dust_mass}. In fact, the high temperatures obtained by fitting only a single component to the SED could indicate that we are tracing only the warmer (brighter) phase of dust, and a second component would be required to characterize the colder (fainter) phase. In this sense, we point out that these temperatures (and masses) are luminosity-weighted. Unfortunately, the data coverage of the SED does not allow us to properly constrain a double-component MBB. According to \cite{santini_2014}, the attempt of reproducing both the Wien and the Rayleigh-Jeans side of the MBB spectrum produces an overestimate of the dust temperature, and in turn an underestimate of the dust mass. However, our results are in agreement with \cite{rybak_2019} who measured a $M_d = 4.1^{+0.5}_{-0.6} \times 10^{8}~ M_\odot$ for XID 42 by means of SED fitting, and with \cite{gilli_2014} who derived for XID 539 a $M_d = 4.9 \pm 0.7 \times 10^{8}~ M_\odot$ on the basis of the continuum emission at 1.3 mm. We stress that the dust temperatures do not affect the gas masses in our treatment, since we derive them in the RJ regime (see Sec. \ref{subsec:gas_mass} for the discussion).

\begin{table}
\caption{\label{tab:dust_mass} Dust temperature and mass.
}
\centering
%\resizebox{\hsize}{!}{
\begin{tabular}{cccc}
\hline \hline
XID & T & $M_d$ \\ %& $\nu_{rest}$ (GHz)
   & [K] & [$10^8~M_\odot$] \\
\hline
42   & 65 & $4.0 \pm 0.5$ \\ %& 142.7
337  & 71 & < 1.0         \\ %& 138.7
539  & 65 & $4.8 \pm 0.5$ \\ %& 132.7
551  & 80 & < 0.9         \\ %& 138.7
666 & 69 & $4.2 \pm 0.5$ \\ % & 136.3
746 & 65 & < 1.5         \\ %& 132.1 

\hline    	  
\end{tabular}
%}
\tablefoot{Dust temperature T (derived from the SED fitting, with a fiducial associated error of $\approx \pm 5$ K) and mass of the dust $M_d$ for each XID source. The upper limits on the masses are given at 3$\sigma$ level.
}
\end{table}

\subsection{Gas mass}

\label{subsec:gas_mass}

The mass of the molecular hydrogen (expressed in $M_{\odot}$) in distant galaxies is usually calculated as $M_{H_2}=\alpha_{CO}~L'_{1-0}$, where $L'_{1-0}$ is the luminosity of the CO(1--0) transition and $\alpha_{CO}$ is the assumed conversion factor in units (omitted in the following) of  $\mathrm{~M_{\odot}~(K~km~s^{-1}~pc^2)^{-1}}$. \cite{carilli_2013} suggest a standard value 0.8 \citep{solomon_2005} usually assumed for the star forming nuclei of classical SMGs, and $\sim$4 for normal star forming galaxies (SFGs), accordingly to what is found for the giant molecular clouds (GMCs) in the Milky Way by \cite{bolatto_2013}. The low value assumed for SMGs would imply that there is more CO emission per unit molecular gas mass. \cite{papadopoulos_2012} suggest that this is likely due to the high SFR that induces heating processes, resulting in an increase of the temperature and excitation level of the molecular gas with respect to normal galaxies.
However, the average excitation level in SMGs may be less extreme than in quasars \citep{weiss_2007}, likely due to the fact that in some SMGs the star formation occurs on a larger scale than in compact quasar hosts. Recent works based on the dust and CO comparison showed how a Galactic $\alpha_{CO}$ may hold also for sources up to almost an order of magnitude above the main sequence \citep{genzel_2015,tacconi_2018}.

As for the galaxies in our sample, they are characterized by extreme compactness (i.e., density), a SFR of several hundreds (up to 1600 for XID 42) $\mathrm{M_{\odot}}$/yr, and powerful nuclear activity ($L_{2-10~\mathrm{KeV}}$ up to tens $10^{44}~\mathrm{erg~s^{-1}}$) that contributes up $\sim$15\% of the total FIR (8-1000 $\mu$m) luminosity \citep{circosta_2019}, that likely boosts the excitation level of the gas, implying a low value of $\alpha_{CO}$. \cite{magdis_2012} derived an estimate of $\alpha_{CO}$ for a few samples of both (local and distant) main sequence and starburst galaxies: they derived the molecular mass exploiting the observed $M_{gas}/M_{dust} \propto Z$ relation (where $Z$ is the metallicity), and then obtained $\alpha_{CO}$ measuring $L'_{1-0}$. For local ULIRGs and high-z SMGs they found $\alpha_{CO} \sim 1$. Using the \cite{magdis_2012} relation, for one of our sources \cite[XID 539, which features nearly solar metallicity as shown by][]{nagao_2012} \cite{gilli_2014} constrained the $\alpha_{CO}$ to a value of 0.8-1 from the measured $M_{dust}$ (which is consistent within 1$\sigma$ with $M_{dust}$ derived in this work), in agreement with the standard $\alpha_{CO}=0.8$ suggested by \cite{carilli_2013}. We note that $\alpha_{CO}\approx 1$ is also adopted in all past works on our targets: \cite{coppin_2010} and \cite{gilli_2014} for XID 403, \cite{calistro_2018} for XID 42, and \cite{circosta_2019} for the entire sample. In particular, we to compare the gas masses and column densities with those derived by \citeauthor{circosta_2019} (\citeyear{circosta_2019}, assumed $\alpha_{CO}=0.8$). In addition, \cite{talia_2018} derived the gas mass of the GMASS 0953 from the CO(3--2) and CO(6--5) transitions, adopting $\alpha_{CO} = 0.8$ (motivated by the compactness and the high SFR of the object), finding a good agreement with the gas mass derived from the [\ion{C}{i}] emission; we note that GMASS 0953 is a z$\sim$2.2 SFG very similar to those in our sample, hosting a heavily obscured AGN ($\log N_H > 24$) and with gas size $\sim$1 kpc, $M_{\ast}\sim$ $10^{11}~\mathrm{M_{\odot}}$ and SFR$\sim$210 $\mathrm{M_{\odot}}$/yr).
Given the aforementioned premises, we adopted $\alpha_{CO}=0.8$ to obtain the molecular masses used in Sec. \ref{subsec:col_dens} to derive the column densities. Then, because of the existing uncertainties in the CO luminosity to gas mass conversion factors, we also calculate the molecular gas masses assuming a Galactic $\alpha_{CO}=6.5$ \citep[as in][]{scoville_2016}. This means that both the masses and column densities simply increase by a factor of 6.5/0.8 $\approx$ 8.1, since all the relations are linear.  Assuming $\alpha_{CO}$ > 0.8 then strengthens the hypothesis that the ISM in the host galaxy contributes significantly to the AGN obscuration.

We calculate the gas masses from both the line and continuum emission, to test whether they provide consistent results and address the large uncertainty on the involved conversion factors. We firstly adopt $\alpha_{CO}$ = 0.8 to obtain $M_{H_2}$ from $L'_{1-0}$; we derive $L'_{1-0}$ from the detected high-J CO lines using different CO Spectral Line Energy Distribution (SLED). As for the continuum emission, we derive $L_{850\mu m}$ (rest-frame luminosity at 850 $\mu$m, that is $\approx 353$ GHz) from the observed flux density $S_{obs}$ (at $\lambda \approx 2.1$ mm) adopting a MBB model, and then we exploit the empirical $ L'_{1-0} \propto L_{850\mu m}$ relation from \cite{scoville_2016} to obtain the CO(1--0) luminosity. We note that this is a pure luminosity relation that does \emph{not} imply any assumption on the $\alpha_{CO}$ factor. Similar trends between (both monochromatic and integrated) FIR and CO luminosity are often found in the literature, showing that they are either nonlinear \citep[e.g.,][]{saintonge_2018} or even point to different relations for SFGs and ULIRGs/SMGs (\citeauthor{daddi_2010} \citeyear{daddi_2010}; \citeauthor{genzel_2010} \citeyear{genzel_2010}; see also \citeauthor{carilli_2013} \citeyear{carilli_2013}). These results suggest that assuming a single constant $L'_{CO}/L_{FIR}$ ratio might be too simplistic. %\citep[e.g.,][]{daddi_2010,genzel_2010,carilli_2013,saintonge_2018}, mostly showing how a single law for SFGs, ULIRGs and SMGs \citep[as assumed by][]{scoville_2016} could be simplistic, and at least two power-laws are required to properly fit objects with normal and high SFR.
However, all these relations show a large scatter (at least $\sim$1 dex for the FIR luminosity); for simplicity, we then assume the single functional form $L'_{1-0} \propto L_{850\mu m}$ from \cite{scoville_2016}, also to compare our results with those of \cite{circosta_2019}, who have assumed the same relation. \cite{scoville_2016} then adopt $\alpha_{CO}$ = 6.5 to calibrate directly the $ M_{H_2} \propto L_{850\mu m}$ relation exploiting both the $L'_{1-0}$ measurements and $L_{850\mu m}$ derived from SPIRE and SCUBA fluxes, finding a ratio $\alpha_{850\mu m}$ = $L_{850\mu m}/M_{H_2}$\footnote{We note that while $\alpha_{CO}$ is defined as the ratio between the molecular mass and the CO(1--0) luminosity, $\alpha_{850\mu m}$ is defined as the ratio between the luminosity and the mass.} = $6.7 \times 10^{19}$ $\mathrm{M_{\odot}^{-1}}$ $\mathrm{erg}$ $\mathrm{s^{-1}}$ $\mathrm{Hz^{-1}}$. They chose a single value for all galaxies in their samples (i.e., SFGs, local ULIRGs and SMGs) arguing that a lower factor would be inappropriate for a globally distributed molecular gas. However, they point out that a $\alpha_{CO}$ value several times smaller is often assumed for ULIRGs and SMGs, and that such a low $\alpha_{CO}$ would be more appropriate to describe the very nuclear regions. In addition, in \cite{kaasinen_2019}, where the \cite{scoville_2016} relation is exploited to test the use of long-wavelength dust emission as a gas mass tracer, the authors conclude that the main contribution to the uncertainty in the derived molecular masses (from either CO(1--0) or dust continuum) remains the $\alpha_{CO}$ factor. Given this large uncertainty, we additionally derive the gas masses assuming the same Galactic $\alpha_{CO}$ of \cite{scoville_2016}, in order to provide a range of masses that fairly takes this uncertainty into account. We calculate the gas masses from the detected lines assuming $\alpha_{CO}$ = 6.5 and then from the continuum using $\alpha_{850\mu m}$, and finally compare them.

We firstly estimate the molecular mass from the luminosity of the detected lines. We calculate the line flux $F_{CO}$ as the flux density  (derived from the uv-plane fitting, Table \ref{tab:results}) times the line FWHM reported in Table \ref{tab:spectra_fit}. For XID 666 we found a double--peaked line, thus we calculate the flux multiplying the flux density by the sum of the two line widths derived from the two Gaussian components of the fit, which is $636 \pm 72$ km/s. For the undetected sources we consider an upper limit to the flux density equal to three times the rms measured in the peak channel and as line width we assumed a typical value of 400 km/s \cite[e.g][]{carilli_2013}, consistent within 1$\sigma$ with the two single-peaked detected lines. We report $F_{CO}$ in column 2 of Table \ref{tab:bigtab}. We estimate the CO luminosity $L'_{CO}$ in $\mathrm{K ~km ~s^{-1} ~pc^2}$ from  $F_{CO}$, given the redshift, by means of the well known relation reported in \cite{solomon_1992}: $L'_{CO} = 3.25 \cdot 10^{7}~\nu_{obs}^{-2}~(1+\mathrm{z})^{-3}~D_L^2~ F_{CO}$ where $\nu_{obs}$ is the observed frequency (in GHz) and $D_L$ is the luminosity distance (in Mpc).

The luminosity of the observed transition has been converted into $L'_{1-0}$ through the line ratio derived from a CO--SLED. This ratio depends on the excitation level of the molecular gas, that can significantly differ for the various astronomical objects \cite[e.g.,][]{carilli_2013}. Historically, many sources have been classified as QSOs while others as SMGs, mainly because of the different selection techniques used; however, there is a significant overlap in their properties \citep{carilli_2013}. Despite the main sequence nature of the galaxies in our sample, recent works \citep{pozzi_2017,mingozzi_2018} have shown how the AGN contribution to the high-J CO transitions is not negligible even in low-luminosity QSOs ($L_{2-10~\mathrm{keV}}\sim 10^{42-43} \mathrm{erg~s^{-1}}$, for comparison our sources have $L_{2-10~\mathrm{keV}}\sim 10^{44} \mathrm{erg~s^{-1}}$). In addition, in compact high-z galaxies the AGN boost to the high-J transitions can affect the whole source \citep{vallini_2019}, making its effect more important than in lower redshift objects. We also stress that the Mid-IR (5-15 $\mu$m) emission of our sources is by far (> 50\%) dominated by the AGN component (as shown in the SED decomposition by \citeauthor{circosta_2019}, \citeyear{circosta_2019}). Thus, we argue that a main-sequence CO-SLED could not correctly fit the physical properties of the gas in these objects, due to their compactness and high AGN luminosity; however, \cite{daddi_2015} showed that, in a sample of z=1.5 normal SFGs, the high-J emission (at least from the CO(5--4) transition) is likely more similar to that of SMGs, rather than local SFGs (while for low-J transitions -- up to CO(3--2) -- they show a MW-like CO-SLED). At any rate, we point out that assuming a lower excitation CO ladder that SMG would result in an increase of the gas mass and, as a consequence, in a higher contribution of the host galaxy ISM to the AGN obscuration. Given these premises, we use two different AGN CO--SLEDs suitable for QSOs and SMGs, in order to encompass a large range of possible values. This, coupled with the additional continuum-based estimate, partly mediates the large uncertainties in measuring the ISM masses from high-J CO transitions, primarily due to the assumed excitation ladders.
We use ratios between the observed and CO(1--0) lines reported in Table 2 of \cite{carilli_2013}, who derived averaged values for different selected object types (including QSO and SMG) based on all the available literature at the time. The CO(7--6)/CO(1--0) ratio (needed for XID 539) and CO(6--5)/CO(1--0) ratio (needed for XID 337 and XID 551) are not reported, so we have extrapolated them from the median of the CO--SLEDs presented in the same work. They are equal to 0.16 and 0.48 (for SMG and QSO, respectively) for the CO(7--6)/CO(1--0) ratio, and equal to 0.34 and 0.64 for the CO(6--5)/CO(1--0) ratio. These values for both the transitions are in good agreement (within 1$\sigma$) with those reported by \citeauthor{kirkpatrick_2019} (\citeyear{kirkpatrick_2019}, see Table 4) for AGN-dominated galaxies. We assume in the following the same designation of \cite{carilli_2013} for the two adopted conversion factors, namely SMG CO--SLED and QSO CO--SLED.

We also estimated the molecular gas mass from the continuum luminosity. We firstly exploit the empirical relation between the luminosity of the CO(1--0) transition $L'_{1-0}$ and $L_{850\mu m}$ (in units of $ \mathrm{erg~s^{-1}~Hz^{-1}}$) shown in the left panel of Fig.1 in \cite{scoville_2016}, and valid for normal star forming galaxies, local ULIRGs, and SMGs: $L'_{1-0} = 3.02\times 10^{-21} ~L_{850\mu m}~\mathrm{K ~km ~s^{-1} ~pc^2}$. Given the luminosity distance $D_L$ of a source, $L_{850\mu m}$ can be derived from the red-shifted flux density at 850 $\mu$m $S_{\frac{\nu_{850\mu m}}{1+z}}$ as $4 \pi D_L^2 S_{\frac{\nu_{850\mu m}}{1+z}}\mathrm{(1+z)^{-1}}$.
Thus, we firstly need to rescale the measured flux densities $S_{obs}$ to $S_{\frac{\nu_{850\mu m}}{1+z}}$. We assume a MBB law in the optically thin regime $S_\nu \propto B_\nu(T_{d}) (1-e^{-\tau}) \approx B_\nu(T_{d})\tau_\nu$, where $\tau_\nu=(\nu/\nu_0)^\beta$ is the dust optical depth, with $\beta=2$ and $\nu_0=1.5$ THz, defined as the frequency at which the dust becomes optically thick \cite[e.g.,][]{conley_2011, rangwala_2011, gilli_2014}; at this wavelength the Planck function can be approximated by the Rayleigh--Jeans law $B_{\nu,RJ}(T_{d}) = 2 k_B T_{d} \nu^2 /c^2$, thus $S_\nu \propto \nu^{\beta+2}$. Hence, from the ratio between $S_{\frac{\nu_{850\mu m}}{1+z}} \propto [353~\mathrm{GHz}/(\mathrm{1+z})]^{\beta+2}$ and $S_{obs} \propto \nu_{obs}^{\beta+2}$ we derive
$S_{\frac{\nu_{850\mu m}}{1+z}} = S_{obs}~(353~\mathrm{GHz}/\nu_{obs})^{\beta+2} (\mathrm{1+z})^{-({\beta+2})}$.
These flux densities are reported in the fourth column of Table \ref{tab:bigtab}. We compared these values with those obtained from both the template \cite[performed by][]{circosta_2019} and MBB ($\beta$=2, Sec. \ref{subsec:dust_mass} of this work) SED fitting.
%In addition, \cite{circosta_2019} also performed a SED fitting using a MBB ($\beta$=2), in order to derive the $L_{850\mu m}$ and the molecular masses through the \cite{scoville_2016} empirical relation. The reported relative error on these luminosity is 20\%, which is then assumed for the flux densities derived by the MBB SED fitting performed in this work in Sec. \ref{subsec:dust_mass}.
In general, the fluxes from the template fitting are a factor $\sim$1.7 (XID 539)- 4.9 (XID 746) higher than those derived from the MBB SED fitting. This could be attributed to the intrinsic differences between the models, to the dependence of the flux density on the RJ slope (i.e., the assumed $\beta$) and to the uncertainties due to the lack of data in the RJ tail. In this respect, the MBB SED fitting in Sec. \ref{subsec:dust_mass} does include the additional points (or upper limits) at 2.1mm of this work. We found that the MBB SED fitting derived flux densities are in excellent agreement (~1$\sigma$) we those derived in this section and reported in Table \ref{tab:bigtab}.
%In addition, note that in \cite{circosta_2019} the temperatures used to calculate the $L_{850\mu m}$ exploited in the \cite{scoville_2016} empirical relation were calculated performing a separate SED fitting using the same MBB, $\beta$=2 model, so that the derived gas masses can be fairly compared to those of this work.

Finally, we can write the 850 $\mu$m luminosity as:

$$L_{850\mu m} = 4 \pi D_L^2 \left( \frac{353~\mathrm{GHz}}{\nu_{obs}} \right)^{\beta+2} S_{obs}~\mathrm{(1+z)^{-({\beta+3})}}~.$$

We note that this luminosity, as well as the derived gas masses and column densities, does not depend on the dust temperature, due to the RJ regime assumption. In order to obtain the total gas mass we also need to account for the atomic hydrogen mass $M_{HI}$. We assume a typical ratio $M_{H_2}/M_{HI}\approx5$ as found by \cite{calura_2014} on the basis of IR observations of a large sample of high-redshift QSOs (z > 5.7).

\begin{figure}
\resizebox{\hsize}{!}{\includegraphics{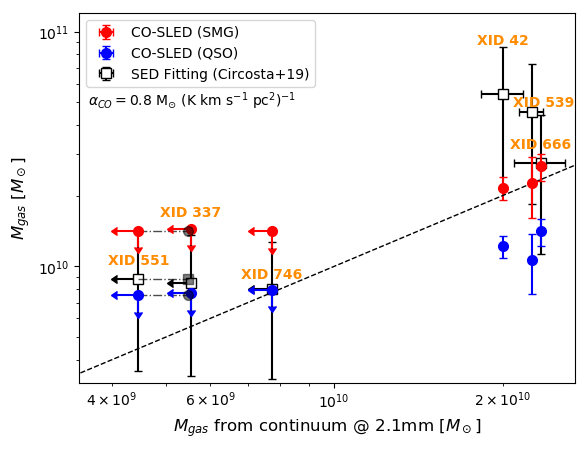}}
\caption{Gas mass derived from the dust emission (x-axis) vs. gas mass derived from the CO line (y-axis). We used $\alpha_{CO}$=0.8 to calculate the molecular gas mass from both the line and continuum emission, which have been previously converted into $L'_{1-0}$ assuming different CO--SLEDs (SMG, red filled circles and QSO, blue filled circles) for the line, and the $ L'_{1-0} \propto L_{850\mu m}$ empirical relation from \citeauthor{scoville_2016} (\citeyear{scoville_2016}, see left panel of their Fig. 1) for the continuum. We accounted for the atomic mass assuming it is 1/5 of the molecular mass. We also report on the y-axis (black open squares) the gas mass derived from the SED fitting by \cite{circosta_2019}. The source IDs are reported in orange. For clarity, we slightly shifted the datapoints of XID 551 toward the left, as they would overlap with those of XID 337 (black shaded symbols). In addition, we report the horizontal error bars of the gas mass derived from the dust emission only on the black open squares. The black dashed line represents the 1:1 relation.}
\label{fig:M_M}
\end{figure}

In Table \ref{tab:bigtab} we report the luminosities of the observed transitions (column 3), the CO(1--0) luminosities (column 6) and gas masses (column 7) derived from both the CO--SLEDs and continuum (as indicated in column 5). In Fig. \ref{fig:M_M} we show the gas masses derived from the detected CO lines, assuming the two different CO--SLEDs, vs. the gas masses derived from the dust continuum emission. We also report the masses obtained by \cite{circosta_2019} from the SED fitting performed in the infrared band.% \textbf{using the MBB ($\beta$=2) model}.

For the detected sources (XID 42, XID 539 and XID 666) the gas masses obtained from the dust emission are fully compatible (within 1$\sigma$) with those derived from both the SMG CO--SLED conversion factor and IR-SED fitting \citep{circosta_2019}. Moreover, for XID 42, \cite{calistro_2018} obtained a gas mass of 5 $\pm$ 2 $\times 10^{10} M_\odot$ from the CO(3--2) line, which is in agreement with our SMG CO--SLED and continuum measurements (3$\sigma$) and SED fitting derived mass (1$\sigma$).
XID 539 shows all masses compatible within 3$\sigma$ with the molecular mass of $\simeq$ 1.6 $\times 10^{10} M_\odot$ found by \cite{coppin_2010} on the basis of CO(2--1) transition analysis.
The lowest masses are the those derived using the QSO CO--SLED, as a consequence of the higher line ratios \citep{carilli_2013}, which in turn lead to lower $L'_{1-0}$ values.
Finally we note that for the undetected sources the gas masses derived by \cite{circosta_2019} from the SED fitting are in agreement (within 1$\sigma$) with the upper limits derived in this work.

\begin{figure}
\resizebox{\hsize}{!}{\includegraphics{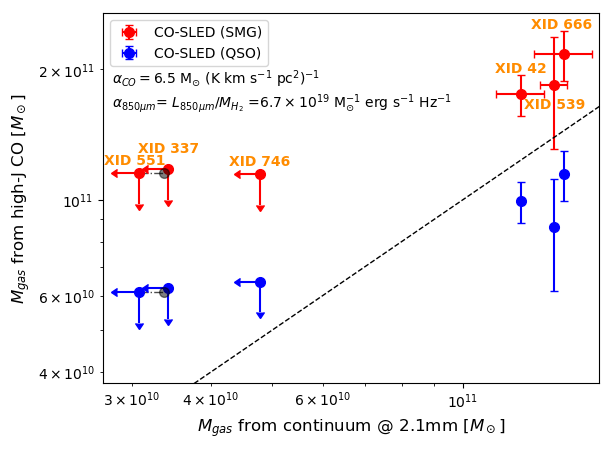}}
\caption{Gas mass derived from the dust emission (x-axis) vs. gas mass derived from the CO line (y-axis). We used $\alpha_{CO}$=6.5 to calculate the molecular gas mass from the line, which has been previously converted into $L'_{1-0}$ assuming different CO--SLEDs (SMG, red filled circles and QSO, blue filled circles), and $\alpha_{850\mu m}$ from \citeauthor{scoville_2016} (\citeyear{scoville_2016}, see right panel of their Fig. 1) to derive the molecular gas mass from the continuum. We accounted for the atomic mass assuming it is 1/5 of the molecular mass. The source IDs are reported in orange. For clarity, we slightly shifted the datapoints of XID 551 toward the left, as they would overlap with those of XID 337 (black shaded symbols). In addition, we report the horizontal error bars of the gas mass derived from the dust emission only on the black open squares. The black dashed line represents the 1:1 relation.}

%\caption{Gas mass derived from the dust emission (x-axis) vs. gas mass derived from the CO line (y-axis). On the plot we report the $\alpha_{CO}$ used to calculate the molecular mass from the line, which has been previously converted into $L'_{1-0}$ assuming different CO--SLEDs (SMG, red filled circles and QSO, blue filled circles) and the $\alpha_{850\mu m}$ from \cite{scoville_2016} used to calculate the molecular mass from the continuum. We accounted for the atomic mass assuming it is 1/5 of the molecular mass. The source IDs are reported in orange. For clarity, since the points of XID 337 and XID 551 are almost superimposed, for the latter we shifted the real points (black shaded symbols) toward the left. In addition, we report the horizontal error bars of the gas mass derived from the dust emission only on the red filled circles. The black dashed line represents the 1:1 relation.}
\label{fig:M_M_sco}
\end{figure}

In order take in account also a Galactic $\alpha_{CO}$, we calculate the gas mass (still accounting for the atomic gas mass as 1/5 of the molecular gas mass) from the CO emission assuming $\alpha_{CO}$ = 6.5, as used by \cite{scoville_2016} to calibrate their relation (this $\alpha_{CO}$ also includes a factor of 1.36 with respect to the molecular hydrogen gas mass, accounting for the mass of heavier elements). Then, we calculate the molecular gas mass directly from $L_{850\mu m}$ via the $\alpha_{850\mu m}$ factor (we note that $\beta=1.8$ assumed by \cite{scoville_2016} to derive the $L_{850\mu m}$ -- and used to calibrate $\alpha_{850\mu m}$ -- slightly differs from that assumed in this work, that is $\beta=2$; this difference is negligible considering the large uncertainty on the other assumed conversion factors). The results are shown in Fig. \ref{fig:M_M_sco}.
Because of the linear scaling of the gas mass with $\alpha_{CO}$ at a given $L'_{1-0}$, the CO-derived gas masses in Fig. \ref{fig:M_M_sco} are a factor 6.5/0.8 = 8.1 times higher than those shown in Fig. \ref{fig:M_M}. For the detected sources, we found that the gas mass derived from the continuum is comprised in the range of values encompassed by the gas masses derived assuming the two CO--SLEDs, and in general that they are all compatible within 3$\sigma$. The consistency between these measurements is further increased if one considers that an additional $\sim$25\% uncertainty is present in the $\alpha_{850\mu m}$ conversion factor.

\subsection{Notes on the morphology of the sources}
\label{subsec:morph}
\begin{figure}[h]
		\centering
		\resizebox{\hsize}{!}{\includegraphics[width=0.9\textwidth,keepaspectratio]{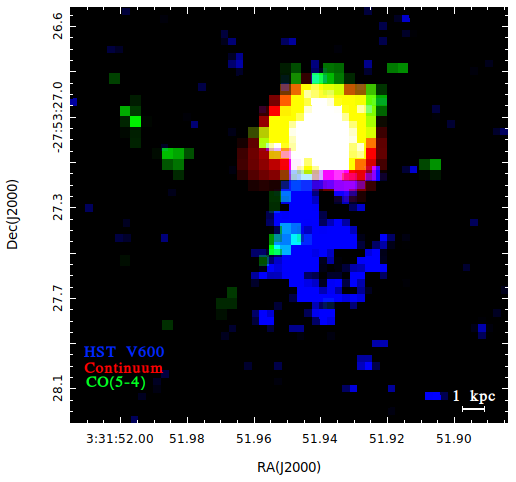}}		
		\caption{Composed r-g-b image of XID 42: the red, green and blue channels are the dust continuum, CO integrated flux and HST $V_{600}$ image, respectively. In the bottom-right corner the physical scale is reported.
		}
		\label{fig:42_HST}
\end{figure}

XID 42 shows a gradient along one direction in the velocity map, at least with a beam separation, suggesting the presence of a differential kinematic structure. In addition, its velocity dispersion is mostly concentrated in the central core of the galaxy. \cite{calistro_2018} have inferred its inclination angle ($40^{\circ} \pm 50^{\circ}$) from the axial ratio of the CO(3--2) emission assuming a disk--like geometry, while \cite{rybak_2019} have fit the  [\ion{C}{ii}] emission with a rotating disk model. We investigated the observation of the source carried out by the HST in the V band (GEMS tile ID: 27), corresponding to the rest--frame UV emission. In Fig. \ref{fig:42_HST} we report the r-g-b image in which the red, green and blue channels correspond to the continuum dust emission, CO moment 0 and optical image, respectively.
Originally the optical image was shifted by $\sim$0.2 arcsec from the center of the dust (and CO) emission in the northwest direction, in agreement with the astrometric error between ALMA and HST images shown by \cite{dunlop_2017}. Thus, we corrected the HST image astrometry using the \textit{GAIA} position measurements of the sources in the same field \citep{gaia_collab}. After the correction the bulks of the stellar, dust and molecular gas emission coincide, within the beam resolution. The optical image shows a very irregular shape that, coupled with the significant shift of the CO(5--4) line with respect to the optical rest--frame (Fig. \ref{fig:spectra}), confirms the peculiar properties of this source. Therefore, we point out that XID 42 may not be securely associated to a regular rotating disk morphology, and further high-resolution observations are required to draw definitive conclusions on its nature.

Indications of a rotating system for XID 666 come from the observed velocity gradient that is clearly distributed along a defined direction (moments 1, Fig. \ref{fig:moments}), and from the line that shows a double--peaked feature (Fig. \ref{fig:spectra}), with the blue and red--shifted peaks that actually correspond to two different emitting regions. However, the very asymmetric distribution of the two peaks suggests the possibility that XID 666 could actually be a blend of two merging systems. As a support to this interpretation, the moment 2 map (Fig. \ref{fig:moments}) shows that the velocity dispersion is mostly concentrated in two clumps, a bigger one located in the innermost region of the source, and a smaller one located southeast from the main clump. Even improving the resolution during the imaging (i.e., lowering the robust weighting), we could not unambiguously distinguish the presence of two different sources, and future observations at higher resolution will help to reject or confirm this interesting prospect.
For XID 539 it would be critical to assess the presence of a rotating system, due to both the spectral shape (Fig. \ref{fig:spectra}) and low S/N per channel (< 5 in the peak). However, the velocity maps suggest the presence of a differential kinematic structure in one direction, at least with a beam separation between the red and blue--shifted region, and the velocity dispersion is confined in a relatively small region concentrated at the center of the source (moments 1 and 2, Fig. \ref{fig:moments}). In addition, \cite{debreuck_2014}, using ALMA data with a resolution of $\sim$0.5", demonstrated that the  [\ion{C}{ii}] emission is well described by a rotating disk model with a similar position angle.\\
Despite these indications of rotating systems, we could not properly constrain a realistic disk-like model for the sources, due to the large uncertainties on the sizes. Instead, in order to derive the column densities in Sec. \ref{subsec:col_dens}, we consider only a simple uniform spherical geometry. However, we point out that assuming a more compact morphology would simply result in compressing the same amount of gas reservoir in a smaller volume, implying higher densities. Future observations at higher resolution will dramatically reduce the uncertainties on the sizes and in turn on all the size-dependent quantities such as the ISM density.

\subsection{Column density associated with the ISM}
\label{subsec:col_dens}

The obtained masses and sizes reveal that a large reservoir of gas is confined in a few kiloparsec region, in agreement with the literature for SMG in the same redshift range \citep[e.g.,][]{carilli_2013,bothwell_2013}. We derived the ISM density as $M_{gas}/V$, where $M_{gas}$ is the gas mass derived either from the CO line (using CO--SLEDs) or the continuum emission (using the \cite{scoville_2016} $L_{850\mu m} \propto L'_{1-0}$ relation), assuming $\alpha_{CO}$ = 0.8 in both the cases, and where V is the volume of a given geometrical model of the source. In order to obtain the number density of the hydrogen atoms, we further divide by the proton mass. Then, we calculated the column density toward the nucleus multiplying the number density by the length of the optical path, from the center of the source toward the galaxy outskirts.

\begin{figure}
\resizebox{\hsize}{!}{\includegraphics{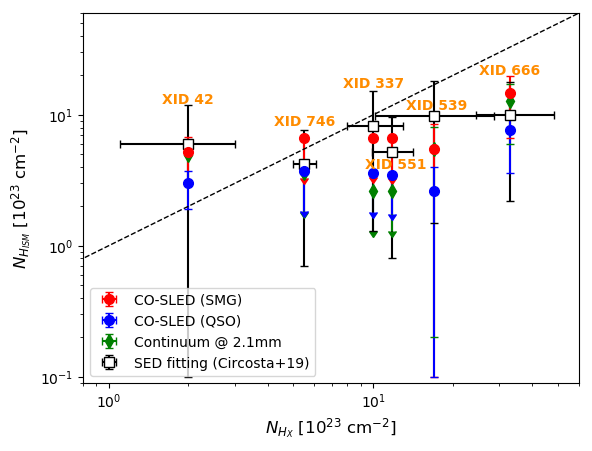}}
\caption{X-ray derived column densities (x-axis) vs. ISM column densities obtained with a uniform spherical model (y-axis). The values related to the SMG CO--SLED, the QSO CO--SLED, and the continuum emission method are indicated by the red circles, blue circles, and green diamonds, respectively. We also report on the y-axis (black open squares) the column densities derived from the SED fitting by \cite{circosta_2019}. The source IDs are reported in orange. The black dashed line represents the $N_{H_X} = N_{H_{ISM}}$ curve. The error bars correspond to the 1$\sigma$ significance level, while the upper limits are given at the 3$\sigma$ level. For clarity the horizontal error bars of the X-ray derived column densities are reported only on the black open squares.}
\label{fig:NH_NH}
\end{figure}

We present a homogeneous spherical model for every source, which is the simplest assumption because it is symmetrical in any direction and independent on the inclination angle. Given the 3$\sigma$ upper limit to the gas mass for the undetected sources, we derived a 3$\sigma$ upper limit on the column densities as well. As the source diameter we considered the mean value between the major and minor axis of the Gaussian ellipse fit reported in the top panel of Table \ref{tab:phys_dim}; as path length we assumed half of the diameter.
In Table \ref{tab:bigtab} we report the gas column densities $N_{H_{ISM}}$ (column 8) for the spherical model using the gas mass derived with the various methods indicated in column 5 of the same table. In column 11 we report the X-ray column density $N_{H_X}$ derived by \cite{circosta_2019}. The ISM vs. X-ray column densities plot is shown in Fig. \ref{fig:NH_NH}.

All the column densities derived through the different methods show values similar to $N_{H_X}$ (up to $10^{24}~\mathrm{cm}^{-2}$), suggesting that the host galaxy can provide a significant amount of the column density toward the central AGN. These results are also in agreement with the column densities found by \cite{circosta_2019} from the masses obtained by the IR-SED fitting. We stress that if we consider the gas masses derived from the high-J CO lines and from the continuum emission using $\alpha_{CO}=6.5$ and $\alpha_{850\mu m}$, respectively, the ISM column density would increase by a factor of 8.1.

\begin{table*}
\caption{\label{tab:bigtab} Summary of the source properties. Column densities refer to the uniform spherical model.
}
\centering
\resizebox{\hsize}{!}{
\begin{tabular}{cccc|cccc|ccc}
\hline\hline

XID      & $\mathbf{F_{CO}}$&  $L'_{\mathrm{CO, obs}}$&$S_{\frac{\nu_{850\mu m}}{1+z}}$ &	  Method	& $L'_{1-0}$	   &  $M_{gas}$ 			  &	 $N_{H}$	   &   $M_{\mathrm{gas, SED}}$   &    $N_{H, SED}$     &	 $N_{H, X}$		 \\
(1)      &	   (2)      &	     (3)	      &    (4)         &   (5)      &	 (6)	       &    (7) 			   &	     (8)	    &	      (9)		  &	 (10)		&    (11)		       \\		 
\hline
            &		    &			      & 		&  SMG CO--SLED      & $ 2.2 \pm 0.2$	& $2.1 \pm 0.2$        &   5.2$^{+1.6}_{-2.2}$  &			    &			  &				 \\
42          &$0.55 \pm 0.06$& $ 0.9 \pm 0.1$	      & $35 \pm 3$	&  QSO CO--SLED      & $1.3 \pm 0.1 $	& $ 1.2 \pm 0.1$       &   3.0$^{+0.9}_{-1.3}$  &      $5.41 \pm 3.21$      &	 $6.0 \pm 5.9$    &    $2.0^{+1.0}_{-0.9}$	\\ 
            &		    &			      & 		&  MBB        & $2.1 \pm 0.2 $   & $2.0 \pm 0.1$     &   $4.9^{+1.4}_{-2.0}$ &  			   &			 &				\\
\hline
            &		    &			      & 		&  SMG CO--SLED      & $ < 1.5 $	& $< 1.4$		  &   $< 6.7$  	   &				 &		       &			      \\
337         & $< 0.32$      & $< 0.5$		      & $< 7$		&  QSO CO--SLED 	  & $< 0.8$	     & $< 0.7$  		 &   $<$ 3.6		  &	$0.85 \pm 0.51$ 	&    $8.2 \pm 6.9$    &     $10.0^{+3.0}_{-2.0}$     \\
            &		    &			      & 		&  MBB        & $< 0.6$ 	 & $< 0.5$		  &   $ < 2.6$  	   &				 &		       &			      \\
\hline
            &		    &			      & 		&  SMG CO--SLED      & $ 2.3 \pm 0.7$	& $2.2 \pm 0.6 $    &	5.5$^{+3.0}_{-5.4}$  &  			 &		       &			      \\
539         &$0.22 \pm 0.06$& $ 0.4 \pm 1 $	      & $17 \pm 1$	&  QSO CO--SLED 	  & $1.1 \pm 0.3 $   & $ 1.0 \pm 0.3$	 &   2.6$^{+1.4}_{-2.5}$  &	 $4.55 \pm 2.70$	 &    $9.8 \pm 8.3$    &     $17.0^{+11.7}_{-8.8}$  \\  
            &		    &			      & 		&  MBB        & $2.3 \pm 0.1$	& $2.2 \pm 0.1$      &   $5.5^{+2.6}_{-5.3}$  & 			    &			  &				 \\
\hline
            &		    &			      & 		&  SMG CO--SLED 	  & $ < 1.5 $	     & $< 1.4  $		   & $< 6.6  $  	    &				  &			&			       \\
551         &$< 0.31$	    & $< 0.5$		      & $< 7$		&  QSO CO--SLED 	  & $< 0.8$	     & $< 0.7  $		   & $<$ 3.5		    &	  $0.88 \pm 0.52$	  &    $5.2 \pm 4.4$	&    $11.8^{+2.4}_{-1.9}$      \\
	    &		    &			      & 		&  MBB        & $< 0.6$ 	 & $ < 0.5 $		   & $ < 2.6$		    &				  &			&			       \\	  
\hline
            &		    &			      & 		&  SMG CO--SLED      & $ 2.8 \pm 0.4$	& $2.7 \pm 0.3$     & 14.6$^{+5.2}_{-7.9}$   &  			 &		       &			      \\
666         &$0.64 \pm 0.08$& $ 1.3 \pm 0.2 $	      & $52 \pm 5$	&  QSO CO--SLED 	  & $1.5 \pm 0.2 $   & $ 1.4 \pm 1.9$	  &  7.7$^{+2.7}_{-4.1}$   &   $2.76 \pm 1.63$    &    $10.0\pm 7.8$	&     $32.8^{+15.4}_{-8.4}$    \\  
            &		    &			      & 		&  MBB        & $2.4 \pm 0.3$	 & $2.3 \pm 0.2$       & $12.8^{+4.4}_{-6.8}$  &			     &  		   &				  \\
\hline     
            &		    &			      & 		&  SMG CO--SLED      & $ < 1.5 $	& $<1.4$			 & $<$ 6.6		  &				&		      & 			     \\ 			       
746         &$ <0.33$	    & $< 0.6$		      &$< 13$	      &  QSO CO--SLED	  &  $< 0.8$	     & $< 0.8 $ 		 & $<$ 3.7		  &  $0.80 \pm 0.47$		&    $4.2 \pm 3.5$    & 	 $5.5^{+0.6}_{-0.5}$  \\
            &		    &			      & 		&  MBB        &  $< 0.8$	 & $ < 0.8 $		    & $<3.6$		    &				  &			&				\\

\hline
\end{tabular}
}
\tablefoot{
Luminosities are expressed in units of $10^{10}~\mathrm{K ~km ~s^{-1} ~pc^2}$, the masses in units of $10^{10} M_{\odot}$ and the column densities in units of $10^{23}~\mathrm{cm}^{-2}$.\\
Column description: (1) ID of the sources. (2) Integrated flux of the observed transition (in units of Jy km $\mathrm{s^{-1}}$), derived multiplying the flux density (second column of Table \ref{tab:results}) by the FWHM of the line (third column in Table \ref{tab:spectra_fit}, for XID 666 we assumed that the line width is equal to the sum of the FWHM of the two line components). For the undetected sources we assumed a line width of 400 km/s.
%For the detected sources it has been derived multiplying the flux density (second column of Table \ref{tab:results}) by the FWHM of the line (third column in Table \ref{tab:spectra_fit}, for XID 666 we assumed that the line width is equal to the sum of the FWHM of the two line components). The upper limits for the undetected sources have been obtained from the 3$\sigma$ upper limit on the flux density and assuming a line width of 400 km/s.}
(3) Luminosity of the observed transition derived from $F_{CO}$. (4) Flux density at 850(1+z) $\mu$m (in units of $\mu$Jy), derived from the observed $\sim$ 2.1 mm flux density assuming a modified black body (MBB) model. (5) Conversion methods adopted to compute the CO(1--0) transition luminosity $L'_{1-0}$: ``SMG CO--SLED'' and ``QSO CO--SLED'' refer to two conversion factors, typical for SMGs and QSOs, respectively, between $L'_{\mathrm{obs}}$ and $L'_{1-0}$ (see text), while ``MBB''exploits the relation $L'_{1-0} = 3.02\times 10^{-21} ~L_{850\mu m}$ found by \cite{scoville_2016}. (6) Derived CO(1--0) transition luminosities. (7) Total gas mass assuming a CO luminosity-to-molecular gas mass ($M_{H_2}$) conversion factor $\alpha_{CO} = 0.8 \mathrm{~M_{\odot}~(K~km~s^{-1}~pc^2)^{-1}}$, and considering the atomic gas mass equal to $\sim M_{H_2}/5$ \citep{calura_2014}. (8) Column densities in the spherical geometry, assuming as a path length of the light the radius of the sphere. (9) (10) (11) Gas mass, infrared SED-fitting column density, and X-ray spectral fitting column density, respectively, derived by \cite{circosta_2019}.\\
The upper limits for the undetected sources are given at the 3$\sigma$ upper level.
}
\end{table*}

%&  $0.22 \pm 0.06$
%&  $0.64 \pm 0.08$
%

The spherical model likely underestimates the source density, since it distributes the mass on a larger volume than models based on more compact and realistic geometries (e.g., disk--like). Given the derived column densities ($\log N_H\sim$23--24) for the spherical model, we infer that a more compact morphology can produce even larger column densities, depending on the inclination angle and gas mass profile. This means that the host galaxy ISM may contribute to the nuclear obscuration up to the Compton-thick regime, especially adopting a high value for $\alpha_{CO}$.

\subsection{Dynamical mass}
Given the indications of a rotating structure for the detected sources (Sec. \ref{subsec:morph}), we derive the dynamical masses in the idealized case of a simple uniform thin disk geometry, where the diameter corresponds to the major axis $a_G$ and the inclination angle $i$ (defined with respect to the perpendicular to the line of sight, north to south) is derived from the axial ratio. We assume $i$ = $32^\circ$, $37^\circ$, $27^\circ$ for XID 42, XID 539 and XID 666, respectively. For XID 42 and XID 539 we derive the dynamical masses $M_{dyn}$ from the FWHM ($\Delta $v) of the CO lines obtained from the spectral fitting reported in Table \ref{tab:spectra_fit}, exploiting the following relation adapted from \cite{wang_2013}, which assumes a ratio of 0.75 between the maximum projected circular velocity $v_{c, proj}$ and $\Delta $v:

$$ M_{dyn} \sin^2 i = 6.5 \cdot 10^4~ \left(\frac{\Delta \mathrm{v}}{\mathrm{km~s^{-1}}}\right)^2 \left( \frac{a_G}{\mathrm{kpc}}\right)~M_\odot~.$$
For XID 666 we resolve the double peak of the line, thus we directly derive the dynamical mass from $v_{c,proj}$ = $195 \pm 14$ km/s, calculated as the half of the peak separation, assuming that $a_G$ is the distance at which the circular velocity flattens:

$$ M_{dyn} \sin^2 i = 1.16 \cdot 10^5~ \left(\frac{v_{c,proj}}{\mathrm{km~s^{-1}}}\right)^2 \left( \frac{a_G}{\mathrm{kpc}}\right)~M_\odot~.$$
We found $M_{dyn} = 3.7_{-0.7}^{+0.8} \times 10^{10}~ M_\odot$, $2.9_{-1.5}^{+1.8} \times 10^{10}~ M_\odot$, $0.8_{-0.3}^{+0.8} \times 10^{10}~ M_\odot$ for XID 42, XID 539 and XID 666, respectively. These values are about one order of magnitude lower than the values previously found for SMG at z$\sim$2 based on the study of resolved and unresolved CO kinematics \citep{tacconi_2008,swinbank_2010,bothwell_2013} and than the baryonic mass calculated as $M_{bar} = M_{\ast}+M_{H_2}+M_{HI}$, where $M_{\ast}$ are the stellar masses obtained by the SED fitting \citep{circosta_2019} which are equal to $M_{\ast} = 2.16 \pm 0.65 \times 10^{11}~ M_\odot$, $2.15 \pm 0.64 \times 10^{11}~ M_\odot$, $4.41 \pm 1.32 \times 10^{11}~ M_\odot$ for XID 42, XID 539 and XID 666, respectively. We stress that, for XID 539, \cite{debreuck_2014} and \cite{coppin_2010} found $M_{dyn}$ in agreement with the value derived in this work. \cite{debreuck_2014} have assumed different inclination angles (29$^{\circ}$, 50$^{\circ}$ and 53$^{\circ}$), while in \cite{coppin_2010} an angle $i=30^{\circ}$ was assumed, as appropriate for randomly inclined disks in a sample of galaxies. 
%The discrepancy between $M_{dyn}$ and $M_{bar}$ could be explained on the basis of our disk model only if a relatively high inclination angle ($|i| \gtsim 80^{\circ}$) is assumed.
Such discrepancy could be attributed either to an overestimate of the stellar mass by the SED fitting or to an underestimate of $\Delta$v and $a_G$, given the sensitivity limitation of the observations. As for XID 666, we assumed that $v_{c, proj}$ is equal to the half of the distance of the peaks and that the disk diameter is the distance at which the velocity profile flattens, but due to the large uncertainty on the source geometry and size it is not possible to securely determine this distance. It is also likely that the stellar mass has been derived considering a different emitting area with respect to the molecular gas. Finally we note that the 30\% error on the stellar masses reported in \cite{circosta_2019} is indicative only of the statistical error given the assumed Salpeter Initial Mass Function \cite[IMF,][]{salpeter_1955}, and the systematic error due to different choices of IMF are likely much larger. By assuming a Chabrier IMF \citep{chabrier_2003} the resulting stellar masses would be a factor $\sim$ 1.7 smaller, leading to a lower discrepancy in the $M_{bar}$/$M_{dyn}$ ratio.

\section{Summary and conclusions}
\label{sec:conclusions}
Exploiting ALMA data we analyzed the high-J CO and the dust emission of a sample of six star forming galaxies hosting a heavily obscured ($\log N_{H_X} > 23$) AGN in the redshift range $\sim$ 2.5 < $z$ < 4.7 in the CDF--S. We detect both the line and continuum emissions for three sources, for which we extracted the flux densities and sizes in order to characterize their masses, gas densities and column densities, and evaluate to what extent the host galaxy ISM can contribute to the central AGN obscuration. Our results are summarized as follows. 

\begin{list}{-}{}

\item We fit the CO line and continuum emission in both the brightness and visibility spaces with a 2--D Gaussian model, finding that all sources are extremely compact objects. The CO emission, tracing the molecular gas component, is contained in regions with major axis in the range 2.1 -- 3 kpc. The continuum emission, tracing the dust component, is produced in somewhat smaller regions, with major axes that are $\sim$ 10 -- 30 \% smaller than those derived from the CO emission. This difference in size can be explained by a radial decrease of the gas column density and dust temperature \citep{calistro_2018}, but we cannot exclude that it could be attributed to the sensitivity limit of the observations.

\item We analyzed the spectra of the detected sources and performed a Gaussian fit to the lines. For one of these sources (XID 666) we found a double--peaked feature that we ascribe to a rotating disk-like morphology. However, given the strong asymmetry of the line shape (Fig. \ref{fig:spectra}, bottom panel) and the dispersion velocity mainly distributed in two separated clumps, we cannot exclude that XID 666 is actually an unresolved merger of two different systems. XID 42 also shows indication of a rotating structure, even if it cannot be securely associated to a regular disk shape: the line is red-shifted by $\sim$500 km/s with respect to the optical rest--frame velocity and the HST image (V--band, Fig. \ref{fig:42_HST}), that traces the UV rest--frame stellar emission, shows a very irregular morphology. These features make XID 42 the most peculiar target among the sample, and observations at higher resolution ($\sim$ 0.1") would be necessary to cover the source with a sufficient ($\sim$ 5-6) number of beams, and unveil whether its kinematics points to relaxed system or to a chaotic, merging, structure.

\item We calculated the gas mass as $M_{H_2}+M_{HI}$, assuming $M_{HI}=M_{H_2}/5$ \citep{calura_2014}. The molecular gas mass was firstly derived from the high-J CO detected lines assuming two different CO--SLEDs typical for SMGs and QSOs to obtain $L'_{1-0}$, and then assuming a $L'_{1-0}$--$M_{H_2}$ conversion factor $\alpha_{CO} = 0.8$. We obtained masses in the range $(0.5 - 2.7) \times 10^{10}~ M_\odot$, in agreement with those derived by \cite{circosta_2019} from the SED fitting. Secondly, we derived the molecular gas masses from the continuum emission on the basis of the $L'_{1-0}$--$L_{850\mu m}$ empirical relation found by \cite{scoville_2016}, using a modified black--body model to calculate $L_{850\mu m}$ and then again $\alpha_{CO}$ = 0.8 to obtain the gas masses. We found that these masses are fully in agreement (within 1$\sigma$) with those derived from the CO line emission. In order to encompass the uncertainty on $\alpha_{CO}$, we also calculated the gas masses from the CO lines using $\alpha_{CO}$ = 6.5, and then compared these values with those obtained from the direct \cite{scoville_2016} relation $M_{H_2}$ and $L_{850\mu m}$, which is based on the assumption of the same $\alpha_{CO}$, finding that they are in good agreement. This corresponds to an increase of the gas masses (a factor of $\sim$8.1) and, as a consequence, in a more pronounced contribution of the ISM to the nuclear obscuration.

%In order to validate also the direct $M_{H_2}$= $L_{850\mu m}$/$\alpha_{850\mu m}$ relation from \cite{scoville_2016}, we recalculated the masses from the CO line using the same $\alpha_{CO} = 6.5$ they used to calibrate it, and then use it to derive the molecular mass from  $L_{850\mu m}$. Finally we compare them, finding that the continuum-derived masses are comprised in the range of values between the gas masses derived assuming the two CO--SLEDs. This suggests that this relation could be reliably used in the redshift range $z$ $\sim 2-4.7$ for this kind of objects.}

\item We investigated the contribution of the host galaxy to the obscuration of the AGN by comparing the column density $N_{H,ISM}$ derived from the ISM masses with that ($N_{H,X}$) derived from the X-ray spectral fitting \citep{circosta_2019}. As for the source geometry, we assumed a uniform spherical model for all the sources. We found that $N_{H,ISM}$ is similar to $N_{H,X}$ ($\sim 10^{23-24}~ \mathrm{cm}^{-2}$), and conclude that the host galaxy can significantly affect the obscuration of the central engine. All the $N_{H,ISM}$ are also comparable to the column densities derived by \cite{circosta_2019} from the IR-SED fitting. We point out that despite the large number of assumptions, the $N_{H,ISM}$ values derived from three independent measurements (i.e., CO line, dust continuum and SED fitting), are in agreement with each other and with $N_{H,X}$. This result strongly supports a scenario in which a substantial fraction of the obscuration may be produced on large (kpc) scales, at least for these FIR-bright objects characterized by a large gas reservoir confined in a very compact volume. Since the spherical model likely underestimates the ISM density, we point out that a more realistic (disk-like) geometry, compressing the gas mass in a smaller volume, would result in larger density; depending on the inclination angle of the disk and the gas distribution, this may produce column densities that can significantly affect the obscuration even in heavily Compton-thick AGN ($\gtrsim 10^{24}~ \mathrm{cm}^{-2}$).

\item From the continuum emission we derived the dust mass, assuming a modified black body law in the optically thin regime. We found that the dust masses are in the range $\sim (0.9-4.9) \times 10^{8}~ M_\odot$.

\item We derived the dynamical masses for the three detected sources, whose values are in the range $(3.7-0.8)\times 10^{10} M_\odot$. These masses are one order of magnitude lower than the sum of the stellar \citep{circosta_2019} and gas masses, and consistent with those derived in previous work for XID 539, assuming different inclination angles \citep{coppin_2010,debreuck_2014}. Such discrepancy could be attributed to an overestimate of the stellar mass by the SED fitting or to an underestimate of the FWHM of the line and the gas component size (given the sensitivity limitation of the observations). Moreover, the stellar mass may be derived considering a different emitting area with respect to the molecular gas. Finally, we note that the scatter of the stellar masses derived by \cite{circosta_2019} does not take into account different assumptions for the IMF. Indeed, by assuming a Chabrier IMF the stellar masses reduce of a factor $\sim$1.7, leading to a smaller discrepancy between $M_{bar}$ and $M_{dyn}$.
\end{list}

From the analysis of our sample, we conclude that SMGs environment is characterized by an extremely dense ISM confined in few kiloparsec. In such conditions, the ISM can produce column densities that contribute to the obscuration of the central AGN up to the Compton-thick regime. In addition, the sources' velocity gradient indicates the presence of a rotating system, even though some of their morphological features suggest a more complex structure, possibly associated with mergers.

\begin{acknowledgements}
     QD acknowledges A. Peca and G. Sabatini for all the support during the long journey, and R. Decarli, M. Mignoli and G. Iorio for the very helpful discussions. CC acknowledges support from the IMPRS on Astrophysics at the LMU (Munich). FV acknowledges financial support from CONICYT and CASSACA through the Fourth call for tenders of the CAS-CONICYT Fund and from CONICYT grants Basal-CATA AFB-170002. This paper makes use of the following ALMA data: 2015.1.01205.S. ALMA is a partnership of ESO (representing its member states), NSF (USA) and NINS (Japan), together with NRC (Canada), MOST and ASIAA (Taiwan), and KASI (Republic of Korea), in cooperation with the Republic of Chile. The Joint ALMA Observatory is operated by ESO, AUI/NRAO and NAOJ.
\end{acknowledgements}

%\appendix

%\section{aacasc}

% WARNING
%-------------------------------------------------------------------
% Please note that we have included the references to the file aa.dem in
% order to compile it, but we ask you to:
%
% - use BibTeX with the regular commands:
   \bibliographystyle{aa} % style aa.bst
  \bibliography{mybib} % your references Yourfile.bib
%
% - join the .bib files when you upload your source files
%-------------------------------------------------------------------

\end{document}